\documentclass[journal,letterpaper,onecolumn,draft,11pt]{IEEEtran}
\usepackage{amsmath}
\usepackage{amsfonts}
\usepackage{mathrsfs}
\usepackage{pifont}
\usepackage{amssymb}
\usepackage{verbatim}
\usepackage{upgreek}
\usepackage{color}
\usepackage[final]{graphicx}
\usepackage{booktabs}
\usepackage{bm}
\usepackage{setspace}
\usepackage{dsfont}
\usepackage{cite}
\usepackage{algorithm}
\usepackage{algorithmic}
\usepackage{epsfig}
\usepackage[english]{babel}
\usepackage[bookmarks=false]{hyperref}
\usepackage{hyperref}
\usepackage{flushend}
\usepackage[]{units}
\usepackage{authblk}
\usepackage{color}
\usepackage{fancyhdr}

\newcommand{\eit}{\end{itemize}}

\newcommand{\ben}{\begin{enumerate}}
\newcommand{\een}{\end{enumerate}}

\newcommand{\bdesc}{\begin{description}}
\newcommand{\edesc}{\end{description}}

\newcommand{\bea}{\begin{array}}
\newcommand{\eea}{\end{array}}

\newcommand{\beqa}{\begin{eqnarray}}
\newcommand{\eeqa}{\end{eqnarray}}

\newcommand{\ds}{\displaystyle}

\newcommand{\Comment}[1]{}

\def\N{{\mathds N}}

\def\R{{\mathds R}}

\def\C{{\mathds C}}

\def\cC{\mbox{$\mathcal C$}}

\def\cH{\mbox{$\mathcal H$}}
\def\cN{\mbox{$\mathcal N$}}
\def\cL{\mbox{$\mathcal L$}}

\def\cX{\mbox{$\mathcal X$}}
\def\cY{\mbox{$\mathcal Y$}}

\newcommand{\be}{\begin{equation}}
\newcommand{\ee}{\end{equation}}

\newcommand{\bzero}{{\mbox{\boldmath $0$}}}

\newcommand{\bA}{\mbox{\boldmath{$A$}}}
\newcommand{\boa}{\mbox{\boldmath{$a$}}}

\newcommand{\bD}{\mbox{\boldmath{$D$}}}

\newcommand{\bI}{\mbox{\boldmath{$I$}}}

\newcommand{\bM}{\mbox{\boldmath{$M$}}}

\newcommand{\bn}{\mbox{\boldmath{$n$}}}
\newcommand{\bbm}{\mbox{\boldmath{$m$}}}
\newcommand{\bP}{\mbox{\boldmath{$P$}}}
\newcommand{\bp}{\mbox{\boldmath{$p$}}}

\newcommand{\bR}{\mbox{\boldmath{$R$}}}
\newcommand{\br}{\mbox{\boldmath{$r$}}}
\newcommand{\bS}{\mbox{\boldmath{$S$}}}

\newcommand{\bU}{\mbox{\boldmath{$U$}}}

\newcommand{\bV}{\mbox{\boldmath{$V$}}}
\newcommand{\bv}{\mbox{\boldmath{$v$}}}
\newcommand{\bW}{\mbox{\boldmath{$W$}}}

\newcommand{\bX}{\mbox{\boldmath{$X$}}}
\newcommand{\bx}{\mbox{\boldmath{$x$}}}

\newcommand{\by}{\mbox{\boldmath{$y$}}}
\newcommand{\bZ}{\mbox{\boldmath{$Z$}}}
\newcommand{\bz}{\mbox{\boldmath{$z$}}}

\newcommand{\balpha}{\mbox{\boldmath{$\alpha$}}}
\newcommand{\bbeta}{\mbox{\boldmath{$\beta$}}}
\newcommand{\btheta}{\mbox{\boldmath{$\theta$}}}

\newcommand{\bgamma}{\mbox{\boldmath{$\gamma$}}}

\newcommand{\bOmega}{\mbox{\boldmath{$\Omega$}}}

\newcommand{\bLambda}{\mbox{\boldmath{$\Lambda$}}}

\newcommand{\dmax}{\begin{displaystyle}\max\end{displaystyle}}
\newcommand{\dmin}{\begin{displaystyle}\min\end{displaystyle}}

\newcommand{\test}{\mbox{$
\begin{array}{c}
\stackrel{ \stackrel{\textstyle H_1}{\textstyle >} }{
\stackrel{\textstyle <}{\textstyle H_0} }
\end{array}
$}}

\newcommand{\testp}{\mbox{$
\begin{array}{c}
\stackrel{ \stackrel{\textstyle H_{1,1}}{\textstyle >} }{
\stackrel{\textstyle <}{\textstyle H_{0,0}} }
\end{array}
$}}

\newcommand{\tr}{\mbox{\rm tr}\, }

\newcommand{\diag}{\mbox{\boldmath\bf diag}\, }

\begin{document}
\title{New ECCM Techniques Against Noise-like and/or Coherent Interferers}

\author{Linjie Yan, Pia Addabbo, \IEEEmembership{Member, IEEE}, Chengpeng Hao, \IEEEmembership{Senior Member, IEEE}, \\Danilo Orlando, \IEEEmembership{Senior Member, IEEE}, and Alfonso Farina, \IEEEmembership{Life Fellow, IEEE}
\thanks{Linjie Yan and Chengpeng Hao are with Institute of Acoustics, Chinese Academy of Sciences, Beijing, China.
E-mail: {\tt yanlinjie16@163.com,haochengp@mail.ioa.ac.cn}. Pia Addabbo is with Universit\`a degli studi Giustino Fortunato, Benevento, Italy. E-mail: {\tt p.addabbo@unifortunato.eu}. D. Orlando is with the Engineering Faculty of Universit\`a degli Studi ``Niccol\`o Cusano'', via Don Carlo Gnocchi 3, 00166 Roma, Italy. E-mail: {\tt danilo.orlando@unicusano.it}. Alfonso Farina is with Selex ES (retired), 00131 Rome, Italy. E-mail: {\tt alfonso.farina@outlook.it}. (Corresponding author: Chengpeng Hao.)}}

\maketitle

\begin{abstract}
Multiple-stage adaptive architectures are conceived to face with the problem of target detection buried
in noise, clutter, and intentional interference.
First, a scenario where the radar system is under the electronic attack of noise-like interferers is considered. In this context,
two sets of training samples are jointly exploited to devise a novel two-step estimation procedure of the
interference covariance matrix. Then, this estimate is plugged in the adaptive matched filter to mitigate the deleterious
effects of the noise-like jammers on radar sensitivity. Besides, a second scenario, 
which extends the former by including the presence of coherent jammers,
is addressed. Specifically, the sparse nature of data is brought to light and the compressive sensing paradigm is applied to estimate
target response and coherent jammers amplitudes. The likelihood ratio test, where the unknown parameters are replaced by previous estimates, is designed
and assessed. Remarkably, the sparse approach allows for echo classification and estimation of both angles of arrival and number of the interfering sources.
The performance analysis, conducted resorting to simulated data, highlights the effectiveness of the newly proposed architectures also
in comparison with suitable competing architectures (when they exist).
\end{abstract}


\section{Introduction}\label{Sec:Intro}
In the last decades, radar art has been significantly influenced by the advances in technology as corroborated by the last-generation
processing boards capable of performing huge amounts of computations in a very short time while keeping the costs relatively low.
This abundance of computation power has allowed for the development of radar systems endowed with more and more sophisticated processing schemes.

To provide a tangible example, let us focus on search radars which are primarily concerned with the detection and tracking of targets
embedded in thermal noise, clutter, and, possibly, intentional interference, also known
as Electronic Countermeasure (ECM) \cite{Richards,antennaBased,EW101,ScheerMelvin}. In this context,
the open literature is rich with novel contributions on
adaptive detection with the result that detection architectures
are evolving towards a continuous performance enhancement.
Consider, for example, the space-time detection algorithms that exploit
large volumes of data from sensor arrays
and/or pulse trains to take advantage of temporal and spatial integration/diversity \cite{kelly1986adaptive,robey1992cfar,gini1,DeMaio-RAO,DeMaioInvCoinc,RicciRao,WLiuRao,GCuiRaoWald,BOR-Morgan,addabboslim,JunLiu00}. Another route followed by the radar 
community to improve the detection performance consists in  using the available information about the structure of the Interference Covariance Matrix (ICM) at the design stage.
As a matter of fact, special structures of the ICM are induced by the system and/or interference properties \cite{Liu1,Liu2,CP00,
JunLiu02,JunLiu01,DeMaioInvPersymmetry,Hongbin1,Cai1992,DeMaioSymmetric,HaoSP_HE,fogliaPHE_SS}.
As an illustration of this fact, consider those decision rules devised assuming that the ICM is centrohermitian.
These algorithms allow us to reduce the number of training samples required for the ICM estimation \cite{GuerciPersymmetry,Liu1,Liu2,Pascal,CP00,
JunLiu02,JunLiu01,DeMaioInvPersymmetry,Hongbin1,Cai1992} by almost a half
while maintaining a satisfactory detection performance.
Further examples are provided in \cite{DeMaioSymmetric,HaoSP_HE}, where it is shown that the spectral symmetry
of the clutter can be used for instance to obtain
gains of about $3$ dB (in SINR, namely Signal-to-Interference plus Noise Ratio), for a Probability of Detection $P_d=0.9$ and Probability of False
Alarm $P_{fa}=10^{-4}$,
in comparison to conventional detectors.

In most of the above contributions, the ICM results from the superposition of two components representative
of the following two interference sources
\begin{itemize}
\item the electronic devices generating thermal noise, which is ubiquitous;
\item the specific operating environment, whose backscattering gives rise to the clutter component, which is
assumed dominant with respect to thermal noise.
\end{itemize}
Additionally, conventional ICM estimation procedures exploit training samples (secondary data) collected in the proximity of the Cell Under Test (CUT).

However, radars might be potential targets of electronic attacks
by an adversary force, which can use, for instance, active techniques aimed at protecting
a platform from being detected and tracked by the radar \cite{ScheerMelvin}. This is accomplished through two approaches: masking and deception.
Noncoherent Jammers or Noise-Like Jammers (NLJs) attempt to mask targets generating nondeceptive interference
which blends into the thermal noise of the radar receiver. As a consequence, the radar sensitivity is degraded due to the increase of the
constant false alarm rate threshold which adapts to the higher level of noise \cite{antennaBased,ScheerMelvin}. In addition,
this increase makes more difficult to discover that jamming is taking place \cite{EW101,FarinaSkolnik}.

On the other hand, the Coherent Jammers (CJs) transmit low-duty cycle signals intended to inject false information into the radar processor.
Specifically, they are capable of receiving, modifying, amplifying, and retransmitting the radar's own signal to create
false targets maintaining radar's range, Doppler, and angle far away from the true position of the platform under
protection \cite{FarinaSkolnik,giniGrecoDRFM,antennaBased,ScheerMelvin}.

Against the aforementioned electronic attacks, radar designers have developed defense strategies referred to as Electronic Counter-CounterMeasure (ECCM)
which can be categorized as antenna-related, transmitter-related, receiver-related, and signal-processing-related depending on the main radar subsystem where
they take place \cite{FarinaSkolnik}.
The first line of defense against jamming is represented by the radar antenna, whose beampattern can be suitably exploited and/or shaped to eliminate
sidelobe false targets or to attenuate the power of NLJs entering from the antenna sidelobes.
The Sidelobe Blanker
(SLB) is an ECCM technique against pulsed interferences \cite{FarinaGini,DeMaioFarinaGini,PiezzoDeMaioFarina} which compares
the detected signal amplitude from the main channel with that of an auxiliary 
channel\footnote{In the following, ``channel'' is used to denote the transmit/receive chain of the radar system \cite{Richards}.}. Specifically, when the auxiliary channel signal power is greater than
that from the main channel, it is likely that the radar is under attack of a CJ from the sidelobes and, hence, 
the detection is blanked. In the presence of
continuous or high duty cycle interferers, the SLB becomes ineffective since it would inhibit the detection of true targets for most of the time.
In these situations, the Sidelobe Canceler (SLC) represents
a viable ECCM against NLJs \cite{antennaBased,FarinaSLC,Reed}. It places nulls in the sidelobes of the main receiver
beam along the directions of arrival of the NLJs which are adaptively estimated using
auxiliary channels. Both the SLB and SLC can be jointly used to face with NLJs and CJs contemporaneously impinging on the sidelobes
of the victim radar \cite{FARINA1995261}.
Finally, it is important to mention that modern radars employ a digitally based approach to implement the SLC function. Specifically, digital samples
from each channel of an electronically scanned array are weighted to adaptively shape the resulting beampattern. These techniques
belong to the more general family of algorithms called Adaptive Digital Beamforming \cite{ScheerMelvin}, which can
be classified as signal-processing-related ECCM.

In this paper, we devise adaptive detection architectures with signal-processing-related ECCM 
capabilities against the attack of NLJs and/or CJs from the
antenna sidelobes.
{
At the design stage, we focus on two operating scenarios which differ for the presence of an unknown number of CJs. 
More precisely,
in the first scenario, the target echoes compete against thermal noise, clutter, and NLJs whose number is unknown, 
whereas the second scenario extends the former by including prospective CJs.
Note that the second scenario is more difficult than the first one, which represents the starting point
for the derivations allowing to easily drive the reader towards the design of more complex systems.
}
Both detection problems are formulated in terms of binary hypothesis tests and, following
the lead of \cite{doubleTraining}, two independent
sets of secondary data are assumed available for estimation purposes.

The first set comes from the conventional radar reference window surrounding the CUT and shares the same ICM components
as the CUT including the clutter component. 
{
The other training set can be acquired by observing that the clutter contribution is, in general, 
range-dependent and tied up to the transmitted waveform. Therefore, it is possible to acquire data free of clutter 
components and affected by the thermal noise and possible jamming signals only.
For instance, for a system employing pulse-to-pulse frequency agility which transmits one pulse,
clutter-free data can be collected before transmitting the pulse waveform by listening to the environment. 
Another example of practical interest concerns radar systems transmitting coherent pulse trains
with a sufficiently high pulse repetition interval. In this case, data 
collected before transmitting the next pulse and at high ranges (or after the instrumental range), 
result free of clutter contribution.
}
However, unlike \cite{doubleTraining}, in this paper, we propose a novel two-step procedure to estimate the ICM components
in a more effective way. Specifically, the thermal noise and NLJ components are estimated using
the second data set\footnote{Note that in \cite{doubleTraining} it is only assumed that the difference between the ICM of the conventional training set
and that of the additional training set is positive semidefinite, while in the present paper information about the structure of
this difference is exploited.} (first step).
The latter estimate replaces the corresponding ICM components of the conventional
data set, which is used to estimate the remaining unknown ICM component, namely, the clutter component (second step).
The number of NLJs impinging on the victim radar is unknown and, hence, is estimated
resorting to either the so-called Model Order Selection (MOS) rules \cite{Stoica_MOS}, which provide more reliable results
than the Maximum Likelihood Approach (MLA) in the presence of nested hypotheses, or a heuristic ad hoc 
procedure based on the MLA.
{ Observe that the last procedure can also be classified as a MOS rule but it does not rely on 
an information criterion as in \cite{Stoica_MOS}.}
More importantly, the herein proposed ICM estimation procedure
requires a less restrictive constraint on the required volume of data with respect to that presented in \cite{doubleTraining} (a point better explained
in Section \ref{Sec:NLJ-only}).
Finally, the detection problem in the presence of NLJs is solved by applying the
two-step Generalized Likelihood Ratio Test (GLRT) design procedure \cite{robey1992cfar} where the ICM of the CUT is replaced
by the new estimate. The final result consists in a multiple-stage architecture capable of taking advantage of the information
carried by the additional training data set.

{
The other considered detection problem also includes the presence of multiple CJs in addition to NLJs, clutter, and 
thermal noise.
Under this assumption, we reformulate the problem at hand in order to bring to light its sparse nature. 
As a consequence, compressive sensing
reconstruction algorithms arise as natural choices to solve it. In the specific case, 
we exploit the Sparse Learning via Iterative Minimization (SLIM) \cite{slim},
due to its trade off between low computational cost and reconstruction performance,
to jointly estimate (under the alternative hypothesis) the unknown target and CJs responses.
}
More precisely, we compute the Likelihood Ratio Test (LRT) where the ICM is replaced
by the previously derived estimate, while target response and CJ amplitudes are estimated by the SLIM.
The exploitation of SLIM (or, generally speaking, compressed sensing algorithms) is due to the fact that, as a byproduct,
it allows for echo classification and estimation of both angles of arrival (AOA) and number of the interfering sources. In fact,
if the LRT statistic is over the detection threshold, the following situations may occur:
\begin{itemize}
\item only CJs are present (target response is zero while CJ amplitudes are nonzero);
\item only the target is present (target response is nonzero while CJ amplitudes are zero);
\item simultaneous presence of the target and CJs (target response and CJ amplitudes are nonzero).
\end{itemize}
With these remarks in mind, we use the estimates provided by SLIM to build up a decision logic capable of discriminating
among the above conditions which, evidently, form a multiple hypothesis test.
Remarkably, this approach can be used in place of the conventional SLB since it recognizes possible CJs echoes
which can be concurrent with target echoes without blanking
the detection. Thus, the proposed detection architecture features SLB/SLC functionalities overcoming the limitations of the SLB.

The remainder of the paper is organized as follows. Section \ref{Sec:ProbForm} is devoted to problem formulation and definition of quantities
used in the next derivations while the design of the detection architectures and estimation procedures are contained in Section \ref{Sec:DetDesign}.
In Section \ref{Sec:Examples}, the behavior of the proposed architectures is investigated by means of numerical examples.
Finally, concluding remarks and future research tracks are given in Section \ref{Sec:Conclusions}. Some derivations are confined in the appendices.

\subsection*{Notation and List of Acronyms}
{ The reader is referred to Table \ref{tab:Acronyms} for the list of the acronyms contained in this paper.}
Moreover, 
vectors and matrices are denoted by boldface lower-case and upper-case letters, respectively.
Symbols $\det(\cdot)$ and $\tr(\cdot)$ denote the determinant and the trace of a square matrix, respectively. Symbols $\bI$ and
$\bzero$ represent the identity matrix and the null vector or matrix of suitable dimensions, respectively. The imaginary unit is denoted by $j$.
Given a vector $\boa$, $\diag(\boa)$ indicates the
diagonal matrix whose $i$th diagonal element is the $i$th entry of $\boa$.
For a finite set $A,\; |A|$ stands for its cardinality.
As to the numerical sets, $\R$ is the set of real numbers,
$\R^{N\times M}$ is the set of $(N\times M)$-dimensional real matrices (or vectors if $M=1$),
$\C$ is the set of complex numbers, and $\C^{N\times M}$ is the set of $(N\times M)$-dimensional complex matrices (or vectors if $M=1$).
The $(k,l)$-entry (or $l$-entry) of a generic matrix $\bA$ (or vector $\boa$) is denoted by $\bA(k,l)$ (or $\boa(l)$).
We use $(\cdot)^T$ and $(\cdot)^\dag$ to denote transpose and conjugate transpose, respectively.
The Clutter-to-Noise Ratio and the Jammer-to-Noise Ratio are denoted by
CNR and JNR, respectively. The conditional probability of an event $A$ given the even $B$ is represented as $P(A|B)$.
Finally, we write $\bx\sim\cC\cN_N(\bbm, \bM)$ if $\bx$ is a complex circular $N$-dimensional normal vector with
mean $\bbm$ and positive definite covariance matrix $\bM$, whereas $\bX=[\bx_1,\ldots,\bx_M]\sim\cC\cN_{N,M}(\bbm, \bM, \bI)$
if $\bx_i\sim\cC\cN_N(\bbm, \bM)$, $\forall i=1,\ldots,M$, and are statistically independent.

\section{Problem Formulation}
\label{Sec:ProbForm}
Consider a radar system which exploits $N$ spatial (identical) channels to sense the surrounding environment. The incoming
signal is conditioned by means of a baseband
down-conversion and a filtering matched to the transmitted pulse waveform.
Next, the output of the matched filter is suitably
sampled and the samples are organized
into $N$-dimensional complex vectors representing the range bins \cite{Richards,BOR-Morgan}.

In what follows, we denote the vector of the returns from the CUT by $\bz\in\C^{N\times 1}$, while the conventional training set,
formed by collecting the returns from the range bins surrounding the CUT \cite{Richards,kelly1986adaptive}, is
stored in the matrix $\bZ=[\bz_1,\ldots,\bz_K]\in\C^{N\times K}$.
Finally, we assume also that the system acquires an additional set of training vectors (free of 
the clutter component and affected by thermal noise and possible NLJs) by listening to the environment
(namely, operating in passive mode) \cite{doubleTraining}. This second set is denoted by $\bR=[\br_1,\ldots,\br_M]\in\C^{N\times M}$.

{ As stated in Section \ref{Sec:Intro}, in this paper we focus our 
attention on two detection problems representative of two scenarios where the latter subsumes the former as a special case.
This choice is dictated by the need to make the derivations easy to be followed. In fact, the scenarios differ
for the presence of CJs in the CUT. }
Specifically, the first problem, which
is the same as in \cite{doubleTraining}, can be formulated as
\be\label{eqn:doubletraining_Problem1}
\left\{
\begin{array}{l}
\begin{aligned}
&H_{1,0}: \left\{
\begin{array}{l}
\bz = \alpha_T \bv(\theta_T) + \bn\\
\bz_{k} = \bn_{k}, \ \br_{m} = \bbm_{m}, \quad k=1,\ldots,K, \ m=1,\ldots,M, \\
\end{array}
\right.\\
&H_{0,0}: \left\{
\begin{array}{l}
\bz = \bn\\
\bz_{k} = \bn_{k}, \ \br_{m} = \bbm_{m}, \quad k=1,\ldots,K, \ m=1,\ldots,M, \\
\end{array}
\right.
\end{aligned}
\end{array}
\right.
\ee
where
\begin{itemize}
\item $\bn$, $\bn_1,\ldots,\bn_K$, $\bbm_1,\ldots,\bbm_M$ are statistically independent random vectors distributed as follows:
$[\bn,\bn_1,\ldots,\bn_K]\sim\cC\cN_{N,K}(\bzero,\bM_1,\bI)$ and $[\bbm_1,\ldots,\bbm_M]\sim\cC\cN_{N,M}(\bzero,\bM_2,\bI)$;
\item $\alpha_T$ is a complex factor representative of the target response and channel effects;
\item $\bv(\theta_T)=\left[1, e^{j 2\pi (d/\lambda) \sin(\theta_T)},\ldots,e^{j 2\pi (d/\lambda)(N-1)\sin(\theta_T)} \right]^T$
is the nominal steering vector with $d$ the array interelement spacing, $\lambda$ the carrier wavelength,
and $\theta_T$ the nominal AOA of the target echoes measured with respect to the array broadside.
\end{itemize}
Unlike \cite{doubleTraining}, we assume that the ICMs exhibit specific structures adhering to situations 
of practical value, namely
$\bM_1 = \sigma^2\bI + \bM_{nj} + \bM_c$ and $\bM_2 = \sigma^2\bI + \bM_{nj}$,
where $\sigma^2\bI$ is the thermal noise component due to the electronic devices with $\sigma^2>0$ the resulting power, $\bM_c$ is
representative of the clutter, and $\bM_{nj}$ is the contribution raising from the presence of NLJs and can be expressed as
$\bM_{nj}=\sum_{i=1}^{N_{nj}} \sigma_{nj,i}^2 \bv(\theta_{nj,i})\bv(\theta_{nj,i})^\dag$
with $N_{nj}$, $\sigma_{nj,i}^2$, and $\theta_{nj,i}$ being the number of NLJs, the power, and the AOA of the $i$th NLJ, respectively.
An important remark on the relationship between the rank of $\bM_{nj}$ and $N_{nj}$ is required for further developments.
{
Precisely, note that when the NLJs are angularly very close to each other, then the inner product
between the resulting NLJ steering vectors is very close to $1$. It follows that the eigendecomposition
of $\bM_{nj}$ leads to a situation where 
the maximum eigenvalue comprises most of the jammers' energy and the associated
eigenvector is representative of the direction from where such energy is transmitted. The remaining
eigenvalues differ by several order of magnitude with respect to the maximum eigenvalue and, hence, they
can be neglected along with the associated eigenvectors. From an alternate point of view, when an orthonormal
basis for the subspace spanned by closely spaced jammer steering vectors is computed by applying 
the Gram-Schmidt process \cite{MatrixAnalysis}, 
it turns out that,
in the new reference system, there exists a dominant component which is several order of magnitude greater than the others.
As a consequence, due to the finite precision of the radar processing unit, the dimension of the subspace
spanned by these steering vectors (and, hence, the rank of $\bM_{nj}$) might be less than or equal 
to the actual number of NLJ steering vectors.
}

The second scenario accounts for the joint presence of NLJs and CJs in the CUT.
This seemingly minor modification leads to a more general and difficult problem, which encompasses the former
and can be written as
\be\label{eqn:doubletraining_Problem2}
\left\{
\begin{array}{l}
H_{1,1}: \left\{
\begin{array}{l}
\ds\bz = \alpha_T \bv(\theta_T) + \sum_{i=1}^{N_{q}}\beta_i \bv(\theta_{q,i}) + \bn,\\
\bz_{k} = \bn_{k}, \ \br_{m} = \bbm_{m}, \quad k=1,\ldots,K, \ m=1,\ldots,M, \\
\end{array}
\right.
\\
H_{0,0}: \left\{
\begin{array}{l}
\bz = \bn,\\
\bz_{k} = \bn_{k}, \ \br_{m} = \bbm_{m}, \quad k=1,\ldots,K, \ m=1,\ldots,M, \\
\end{array}
\right.
\end{array}
\right.
\ee
where $\beta_i$ and $\theta_{q,i}$ are the magnitude and the AOA
of the $i$th CJ, respectively,
$N_q$ is the number of CJs attacking the radar, while the assumptions on $\bn$, $\bn_k$, and $\bbm_m$ keep unaltered. It is clear that problem \eqref{eqn:doubletraining_Problem2} reduces to
\eqref{eqn:doubletraining_Problem1} when $\beta_i=0$, $\forall i=1,\ldots,N_q$.

For future reference, it is worth providing the following definitions. Specifically,
the probability density functions (pdfs) of $\bZ$ and $\bR$ under all the hypotheses
are\footnote{
In what follows, we refer to the ICM of $\bZ$ using the notation $\bM_1$, $\sigma^2\bI+\bM_{nj}+\bM_c$, or $\bM_2+\bM_c$ as well as we refer to
the ICM of $\bR$ writing $\bM_2$ or $\sigma^2\bI+\bM_{nj}$.
}
\be \label{eqn:doubletraining_Problem7}
f(\bZ;\sigma^2,\bM_{nj},\bM_c) = \frac{\exp\{ -\tr[\bM_1^{-1}\bZ\bZ^\dag]  \}}{\left[ \pi^N \det(\bM_1)\right]^K} \quad \mbox{and} \quad
f(\bR;\sigma^2,\bM_{nj}) = \frac{\exp\{ -\tr[\bM_2^{-1}\bR\bR^\dag]  \}}{\left[ \pi^N \det(\bM_2)\right]^M},
\ee
respectively. On the other hand, the pdf of $\bz$ under $H_{l,h}$, $(l,h)\in\{(0,0),(1,0),(1,1)\}$, exhibits the following expression
\begin{multline}
f_{lh}(\bz;l\alpha_T,h \bbeta,h\btheta_q,\sigma^2,\bM_{nj},\bM_c,H_{lh}) = \frac{1}{\pi^N \det(\bM_1)}
\\
\times \exp\left\{ -\tr\left[\bM_1^{-1}\left(\bz-l\alpha_T\bv(\theta_T)-h\sum_{i=1}^{N_{q}}\beta_i \bv(\theta_{q,i})\right)
\left(\bz-l\alpha_T\bv(\theta_T)-h\sum_{i=1}^{N_{q}}\beta_i \bv(\theta_{q,i})\right)^\dag\right]  \right\},
\label{eqn:pdf_z}
\end{multline}
where $\bbeta=[\beta_1,\ldots,\beta_{N_q}]^T$ and $\btheta_q=[\theta_{q,1},\ldots,\theta_{q,N_q}]^T$. Finally, let us denote the
likelihood functions of the distribution parameters as
$\cL_{Z}(\sigma^2,\bM_{nj},\bM_c)=f(\bZ;\sigma^2,\bM_{nj},\bM_c)$, $\cL_{R}(\sigma^2,\bM_{nj})=f(\bR;\sigma^2,\bM_{nj})$,
$\cL_{z}(\alpha_T,\sigma^2,\bM_{nj},\bM_c)=f_{10}(\bz;\alpha_T,0,0,\sigma^2,\bM_{nj},\bM_c,H_{10})$, and
$\cL'_{z}(\alpha_T,\bbeta,\btheta_q,\sigma^2,\bM_{nj},\bM_c)=f_{11}(\bz;\alpha_T,\bbeta,\btheta_q,\sigma^2,\bM_{nj},\bM_c,H_{11})$.

\section{Detection Architecture Designs}
\label{Sec:DetDesign}
In this section, we devise adaptive decision schemes capable of operating under the attack of NLJs and/or CJs. 
In order to simplify the derivations,
we first focus on problem \eqref{eqn:doubletraining_Problem1} where only NLJs are contaminating data and, then, we account for the presence of possible coherent interferers in addition to NLJs.

\subsection{NLJ-only Attack}
\label{Sec:NLJ-only}
{
The design is structured into two parts. In the first part, we present 
an innovative estimation algorithm for $\bM_1$ based upon the MLA
assuming, at the design stage, that the rank of $\bM_{nj}$, $r$ say,
which is representative of the effective interfering sources number, is 
known. The last assumption is motivated by the fact that estimating $r$ through the MLA might 
return erroneous results due to the presence of nested hypotheses. Thus, we first assume that $r$ is known and then 
we replace it with a suitable estimate. To this end, in the second part, we exploit
previous results to conceive multi-stage architectures facing with the situations where $r$ is not known
but bounded from above by the maximum number of NLJs
that is generally known from system specifications and/or the amount of computational resources.
In the detail, we estimate $r$ resorting to Information-based or heuristic ad hoc MOS rules which represent an effective means
to provide reliable estimates of the number of NLJs.
}

Let us focus on problem \eqref{eqn:doubletraining_Problem1} and suppose that the
rank of $\bM_{nj}$ is known\footnote{Recall that the latter might not coincide with
$N_{nj}$ because of the angular separation between the NLJs.}. Unlike \cite{doubleTraining}, the herein proposed estimation procedure exploits all
the available structure information about $\bM_1$ and $\bM_2$. 
{ Since the maximum likelihood estimation of $\bM_1$ and $\bM_2$ through the joint pdf of $\bZ$ and $\bR$ is 
not an easy task at least to best of authors' knowledge, we resort to a two-step suboptimal procedure according to the following rationale
}
\begin{enumerate}
\item use $\bR$ (the additional training set) to find the maximum likelihood estimate (MLE) of $\bM_2$, denoted by
\be
\widehat{\bM}_2=\widehat{\sigma}^2\bI+\widehat{\bM}_{nj}=\underset{\sigma^2,\bM_{nj}}{\arg\dmax} \ \cL_{R}(\sigma^2,\bM_{nj}),
\ee
where $\widehat{\sigma}^2$ and $\widehat{\bM}_{nj}$ are the MLEs of ${\sigma}^2$ and ${\bM}_{nj}$, respectively;
\item compute the MLE of $\bM_c$ based on $\bZ$ assuming that $\bM_2$ is known, namely
\be
\widehat{\bM}_c[\bM_2]=\underset{\bM_{c}}{\arg\dmax} \ \cL_{Z}(\sigma^2,\bM_{nj},\bM_c);
\label{eqn:MLE_Mc}
\ee
\item replace $\bM_2$ in \eqref{eqn:MLE_Mc} with $\widehat{\bM}_2$.
\end{enumerate}
As for the first step, in Appendix \ref{app:M2_estimate}, we show that the expression of $\widehat{\bM}_2$ is
\be
\widehat{\bM}_2=\bU_{S1}{\bD}\bU_{S1}^{\dag},
\label{eqn:M2_estimate}
\ee
where
${\bD}=\textbf{diag}\left\{\frac{\gamma_{1,1}}{M},...,\frac{\gamma_{1,r}}{M},{\sum\limits_{i=r+1}^{N}\gamma_{1,i}}/{M(N-r)},...,
{\sum\limits_{i=r+1}^{N}\gamma_{1,i}}/{M(N-r)}\right\}$
with $\gamma_{1,1}\geq \gamma_{1,2}\geq...\geq\gamma_{1,N}>0$ the eigenvalues of $\bR\bR^\dag$ and $\bU_{S1} \in \C^{N\times N}$ a unitary matrix containing the corresponding eigenvectors. When $r=0$, it is not difficult to show that
${\bD}=\diag\left\{\frac{1}{MN}\sum\limits_{i=1}^{N}\gamma_{1,i},...,\frac{1}{MN}\sum\limits_{i=1}^{N}\gamma_{1,i}\right\}$.


The estimator of $\bM_c$ described in the second step of the procedure is a function of $\bM_2$ which is assumed known.
Thus, the resulting likelihood function depends on $\bM_c$ only and can be recast as
\be
{\cL_{Z}}(\sigma^2,\bM_{nj},\bM_c)={\cL_{Z}}(\bM_c) = \frac{\exp\{ -\tr[(\bM_2+\bM_c)^{-1}\bZ\bZ^\dag]  \}}{\left[ \pi^N \det(\bM_2+\bM_c)\right]^K}.
\label{eqn:L_Zdef}
\ee
In Appendix \ref{app:Mc_estimate}, we prove that the MLE of $\bM_c$ for known $\bM_2$ is given by
\be
\widehat{\bM}_c[\bM_2]=\bM_2^{\frac{1}{2}}\bU_{S2}\widehat{\bOmega}_c\bU_{S2}^{\dag}\bM_2^{\frac{1}{2}}.
\ee
In the last equation, $\widehat{\bOmega}_{c}=\diag\left\{\widehat{\lambda}_{c,1},\ldots,\widehat{\lambda}_{c,N}\right\}$,
where $\widehat{\lambda}_{c,i}=\max\left\{ \frac{{\gamma}_{2,i}}{K}-1, 0 \right\}$, $i=1,\ldots,N$,
with $\gamma_{2,1}\geq...\geq\gamma_{2,N}\geq0$ the eigenvalues of $\bM_2^{-\frac{1}{2}}\bZ\bZ^\dag\bM_2^{-\frac{1}{2}}$ and
$\bU_{S2} \in \C^{N\times N}$ is the unitary matrix of the corresponding eigenvectors.

As final step of the estimation procedure, we replace $\bM_2$ with $\widehat{\bM}_2$ and compute
\be
\widehat{\bM}_1=\widehat{\bM}_2 + \widehat{\bM}_c[\widehat{\bM}_2].
\label{eqn:M1_estimate}
\ee
It is important to observe that this new estimation
procedure (schematically summarized in Algorithm \ref{alg:Framwork}) requires that $M > r$ to ensure that $\widehat{\bM}_2$ is
invertible with probability $1$, instead of $M>N>r$.

\begin{algorithm}[tb!]
\caption{ Estimation Procedure for $\bM_1$ }
\label{alg:Framwork}
\begin{algorithmic}[1]
\REQUIRE $\bR\in\C^{N\times M}$, $\bZ\in\C^{N\times K}$, $r \le N_{nj}$
\ENSURE $\widehat{\bM}_1$
\STATE Compute the Singular Value Decomposition (SVD) of $\bR$ given by $\bR  = \bU_{S1}\bD_R\bV_{R}$
\STATE Compute $\bD_R\bD_R^\dag = \diag(\gamma_{1,1},...,\gamma_{1,N})$
\STATE Compute $\bD=\diag\left\{{\gamma_{1,1}}/{M},...,{\gamma_{1,r}}/{M},\frac{1}{M(N-r)}\sum_{i=r+1}^{N}\gamma_{1,i},...,
\frac{1}{M(N-r)}\sum_{i=r+1}^{N}\gamma_{1,i}\right\}$ or,
when $r=0$, $\bD=\diag\left\{\frac{\sum_{i=1}^{N}\gamma_{1,i}}{MN},...,\frac{\sum_{i=1}^{N}\gamma_{1,i}}{MN}\right\}$
\STATE Compute
$\widehat{\bM}_2=\bU_{S1} \bD \bU_{S1}^{\dag}$
\STATE Compute the SVD of $\widehat{\bM}_2^{-\frac{1}{2}}\bZ$ given by $\widehat{\bM}_2^{-\frac{1}{2}}\bZ = \bU_{S2} \bD_Z \bV_Z^\dag$
\STATE Compute $\bD_Z \bD_Z^\dag = \diag(\gamma_{2,1},...,\gamma_{2,N})$
\STATE Compute $\widehat{\bOmega}_{c}=\diag\left\{\widehat{\lambda}_{c,1},\ldots,\widehat{\lambda}_{c,N}\right\}$, $\widehat{\lambda}_{c,i}=\max\{ {\gamma}_{2,i}/{K}-1, 0 \}, \quad i=1,\ldots,N$
\STATE Compute
$\widehat{\bM}_c[\widehat{\bM}_2]=\widehat{\bM}_2^{\frac{1}{2}}\bU_{S2}\widehat{\bOmega}_c\bU_{S2}^{\dag}\widehat{\bM}_2^{\frac{1}{2}}$
\STATE Return $\widehat{\bM}_1=\widehat{\bM}_2 + \widehat{\bM}_c[\widehat{\bM}_2]$
\end{algorithmic}
\end{algorithm}

Now, we focus on the case where $r$ is unknown and should be somehow estimated from data.
To this end, two different strategies are conceived.

\begin{figure}[htp!]
  \centering
  \includegraphics[scale=0.6]{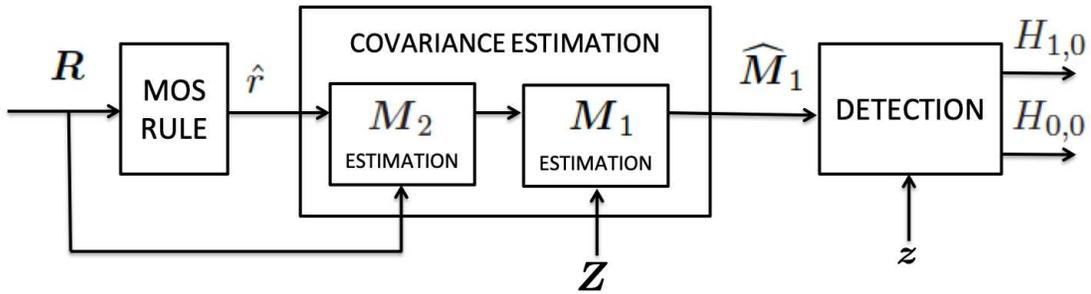}
  \caption{\textit{Three-stage Detection Architectures.}}
  \label{fig1}
\end{figure}

{ The first strategy relies on a three-stage detection architecture (depicted in Figure \ref{fig1}) 
where the first two stages 
(actually, the second stage consists of two sub-blocks)
are devoted to the estimation of $\bM_1$ (and $\bM_2$) and incorporate the Information-based MOS rules \cite{Stoica_MOS}.} More precisely, the first stage provides an estimate
of $r$ and feeds the second stage which is responsible for the estimation of $\bM_2$ and $\bM_1$ according to Algorithm \ref{alg:Framwork}.
The third stage accomplishes the detection task.

The second approach consists
in a modification of the maximum likelihood estimation of $\bM_2$ which accounts for
the significant hop in the order of magnitude of the eigenvalues of $\bR\bR^\dag$ when NLJs are present (a point better explained in Subsection \ref{sec_3a2}).
This discontinuity can be justified by noticing that common JNR values are in the
range $[30,60]$ dB \cite{van1982applied}.
It follows that $r$ can be estimated by detecting this hop in magnitude.

\subsubsection{Three-stage Detection Architectures relying on Information-based MOS Rules}
\label{sec_3a1}
A block scheme of the proposed architectures is depicted in Figure \ref{fig1}: { the first 
two blocks\footnote{Note that the second block is formed by two sub-blocks.} 
perform the estimates of $\bM_2$ and $\bM_1$
exploiting Information-based MOS rules for selecting $r$.} The last block represents the final detection step.
Here, it is important to note that to estimate $r$ the MLA fails because the hypotheses are nested and the likelihood
function monotonically increases with $r$. Thus, focusing on the first block, the estimation of $r$ is accomplished exploiting
the MOS rules which balance the growth of the likelihood
function by means of a penalty term. Specifically, we consider the Akaike Information Criterion (AIC), the Bayesian
Information Criterion (BIC), and the Generalized Information Criterion (GIC) \cite{Stoica_MOS}.
Following the lead of \cite[Ch.7]{VanTrees4}, it is
possible to show that, when $r$ is known, the number of unknown parameters of the distribution of $\bR$ is $k_p(r)=r(2N-r)+1$.
As a consequence, the mentioned MOS rules can be expressed as
\be
\hat{r}=\underset{r\in \{0,...,{N_{nj}^{max}}\}}{\arg\min}\{-2l(\bR,r)+p(r)\},
\ee
where $N_{nj}^{max}$ is the maximum number of jammers and
\be
l(\bR,r)=-MN\log\pi -M \sum_{i=1}^r\log \frac{\gamma_{1,i}}{M}
-M(N-r)\log\left[\frac{1}{M(N-r)}\sum_{i=r+1}^{N}\gamma_{1,i}\right]-MN
\ee
is\footnote{Note that in the case where $r=0$, the term $M\sum_{i=1}^r\log \frac{\gamma_{1,i}}{M}$ does not appear and, in addition, when
$\hat{r}=0$, the procedure returns
\be
\widehat{ M}_2=\diag\left\{ \frac{\sum_{i=1}^{N}\gamma_{1,i}}{MN},\ldots,\frac{\sum_{i=1}^{N}\gamma_{1,i}}{MN} \right\}.
\ee
}
the compressed log-likelihood of $\bR$ assuming that $r$ is known and $p(r)=k_p(r)\nu$ is the penalty term.
Finally, factor $\nu$ takes on the following values
\be
\nu=
\begin{cases}
2,\quad &\textrm{AIC},
\\
1+\rho,\ \rho\geq1, \quad &\textrm{GIC},
\\
\ln M, \quad &\textrm{BIC}.
\end{cases}
\ee
Once the estimate $\hat{r}$ is available, it can be used in place of $r$ in Appendix \ref{app:M2_estimate} to estimate $\bM_2$.
The resulting estimate of $\bM_2$ is, subsequently, used to obtain $\widehat{\bM}_1$ as shown in Appendix \ref{app:Mc_estimate}.

The last block of the proposed architecture implements an adaptive decision rule, devised resorting to the two-step GLRT design
criteria \cite{robey1992cfar}. Specifically, we first compute the
GLRT test assuming that $\bM_1$ is known. Then, the fully adaptive detector is obtained by replacing $\bM_1$ with a suitable estimate.
According to the first step, the GLRT based on the CUT for known $\bM_1$ is the following decision rule
\be
\max\limits_{\alpha_T}\frac{f_{10}(\bz;\alpha_T,0,0,\sigma^2,\bM_{nj},\bM_c)}
{f_{00}(\bz;0,0,0,\sigma^2,\bM_{nj},\bM_c)}\test \eta,
\label{eqn:AMF_01}
\ee
where $f_{l0}(\bz;\ldots)$, $l=0,1$, is defined by \eqref{eqn:pdf_z} and $\eta$ is the detection
threshold\footnote{Hereafter, the generic detection threshold is denoted by $\eta$.}
value to be set according to the desired $P_{fa}$.
It is not difficult to prove that \eqref{eqn:AMF_01} is statistically equivalent to
\be\label{eqn:three stage 1}
\frac{|\bz^{\dag}\bM_1^{-1}\bv(\theta_T)|^2}{\bv^{\dag}(\theta_T)\bM_1^{-1}\bv(\theta_T)}\test \eta.
\ee
Finally, the adaptivity is achieved replacing $\bM_1$ in \eqref{eqn:three stage 1} with the estimate \eqref{eqn:M1_estimate} to come up with
\be
\frac{|\bz^{\dag}\widehat{\bM}_1^{-1}\bv(\theta_T)|^2}{\bv^{\dag}(\theta_T)\widehat{\bM}_1^{-1}\bv(\theta_T)}\test \eta.
\ee
For future reference, we refer to the above decision rule as Improved Double-Trained Adaptive Matched Filter (IDT-AMF), whereas we call the
three-stage architectures coupling the name of the MOS rule used to estimate $r$ and the acronym IDT-AMF. For instance, when BIC is part of the
architecture, we refer to the latter as IDT-AMF-BIC.

\subsubsection{Two-Stage Architecture based upon an Ad Hoc MLE of $\bM_2$}
\label{sec_3a2}
\begin{figure}[htp!]
\begin{center}
\includegraphics[scale=0.6]{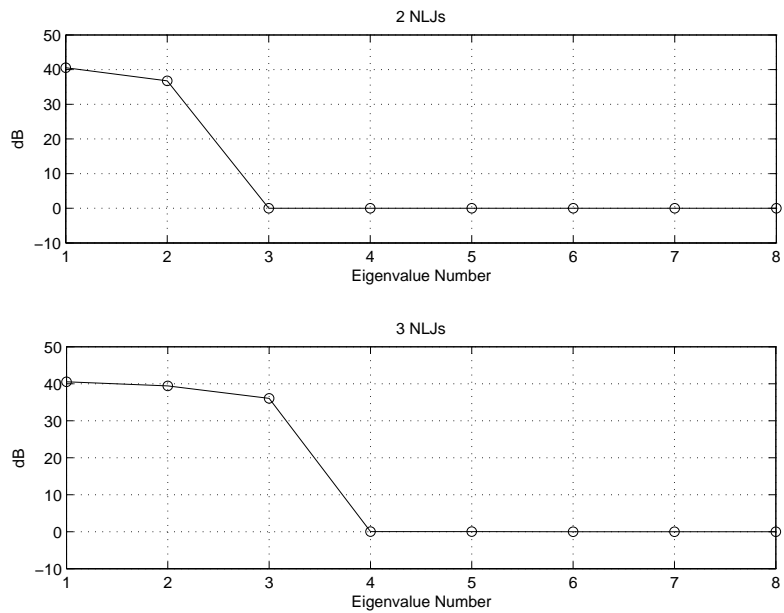}
\caption{Eigenvalues of $\bM_2$ in the presence of NLJs assuming $N=8$ and JNR$=30$ dB for all NLJs.}
\label{fig:dominantEigenvalues2J}
\end{center}
\end{figure}

In the following, we propose a modification of the previously described three-stage architectures 
which consists in removing the block responsible for the estimate of $r$ and incorporating this feature in the block that 
returns the estimates of $\bM_1$ and $\bM_2$ { through a heuristic MOS rule}.
To this end, let us remind that the goal of NLJs is to increase the power noise level within the victim radar making the adaptive threshold as high as possible with
the result of masking the platforms under protection. This fact has some implications for the eigenvalues of the ICM which can be suitably exploited to estimate $r$.
To have a clear vision of this situation, in Figure \ref{fig:dominantEigenvalues2J} we plot the eigenvalues
of $\bM_2$ (which is the true\footnote{In practice, the ICM is estimated from data and quality of the estimate leads to
noise eigenvalue jitter that can be stabilized by means of diagonal loading \cite{ScheerMelvin}.} ICM)
for $N_{nj}=2,3$ NLJs sharing JNR$=30$ dB.
Inspection of the figure highlights that the presence of NLJs breaks down the eigenvalue set
of $\bM_2$ introducing a dramatic drop in magnitude.

The above behavior comes in handy to estimate the model order $r$ by thresholding the difference in magnitude between consecutive eigenvalues of $\bS_2=\frac{1}{M}\bR\bR^{\dag}$,
starting from the lowest values. Specifically, let $\gamma_{1,1}/M\geq\gamma_{1,2}/M\geq...\geq\gamma_{1,N}/M$ be the eigenvalues of $\bS_2$, then
the estimation of $r$ is described in Algorithm \ref{algoXr}, where $\eta$ is a threshold whose value reflects the difference in magnitude
between the eigenvalues associated with both NLJs and thermal
noise and those representative of the thermal noise only.
{
Notice that the underlying decision problem solved 
by this approach is
\be
\begin{cases}
H_1^{''}: \br_m\sim\cC\cN_N(\bzero,\sigma^2\bI+\bM_{nj}), & m=1,\ldots,M,
\\
H_0^{''}: \br_m\sim\cC\cN_N(\bzero,\sigma^2\bI), & m=1,\ldots,M,
\end{cases}
\ee
where the rank of $\bM_{nj}$ is unknown. It turns out that, under $H_0^{''}$, the unknown parameter is
$\sigma^2$, which must be estimated in order to set the detection threshold. To this end, several strategies are possible. 
For instance, a lookup table can be filled up {\em off-line} by measuring the thermal noise power 
under different operating conditions. Then, each entry of this table could be used when the system is in operation.
An alternate approach might consist in scheduling a collection of noisy samples when the antenna is disengaged and 
exploiting such samples to estimate the noise power.
}

Finally, once $r$ has been estimated, $\widehat{\bM}_2$ can be obtained as described in Appendix \ref{app:M2_estimate} and the IDT-AMF is applied.
In the following, we call this architecture Eigenvalue-based IDT-AMF and we use the abbreviation IDT-AMF-EIG.
\begin{algorithm}[htb!]
\caption{ Estimation Procedure for $r$ }
\label{algoXr}
\begin{algorithmic}[1]
\REQUIRE $\eta$, $\gamma_{1,1}/M\geq \gamma_{1,2}/M\geq...\gamma_{1,N}/M>0$
\ENSURE $\widehat{r}$
\STATE Set $i=N-1$, $\hat{r}=0$
\STATE Compute $\Delta_i=\frac{1}{M}(\gamma_{1,i}-\gamma_{1,i+1})$
\STATE If  $\Delta_i>\eta$, then  $\hat{r}=i$ and go to step 6 else go to step 4
\STATE Set $i=i-1$
\STATE If $i\geq1$ go to step 2 else go to step 6
\STATE Return $\widehat{r}$
\end{algorithmic}
\end{algorithm}

\subsection{NLJs and Coherent Interferers Joint Attack}
In this subsection, we focus on problem \eqref{eqn:doubletraining_Problem2} and devise an architecture capable of detecting point-like targets assuming that
noise-like jammers as well as coherent interferers contaminate the echoes from the CUT.
{
Specifically, such architecture consists
of a covariance estimation stage, which relies on the results obtained in Section \ref{sec_3a1}, followed by
a new detection stage which incorporates a sparse reconstruction algorithm. 
This choice is dictated by the fact that problem \eqref{eqn:doubletraining_Problem2} hides an inherent sparse
nature, which can be drawn by means of a suitable reformulation. Thus, we select 
the so-called SLIM algorithm as sparse reconstruction algorithm since it provides a good trade off between computational requirements and reconstruction performance \cite{slim}.
}

\begin{figure}[htp!]
    \centering
    \includegraphics[width=.8\textwidth]{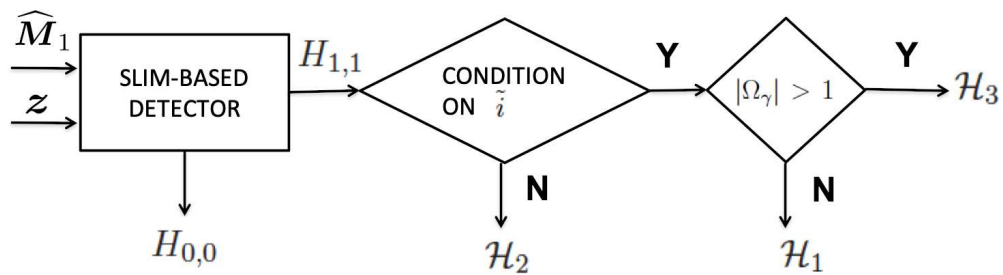}
    \caption{Block scheme of the SLIM-based detection architecture.}
    \label{fig_SLIMarchitecture}
\end{figure}

Let us consider the hypothesis $H_{1,1}$, defined in \eqref{eqn:doubletraining_Problem2}, where it is assumed that a number of coherent
interferers ($N_q$) are present together with the NLJs and note that $\bz$ is the sum of three components
\be
\ds\bz = \alpha_T \bv(\theta_T) + \sum_{i=1}^{N_{q}}\beta_i \bv(\theta_{q,i}) + \bn.
\label{eq_z}
\ee
To effectively apply the SLIM approach, it is necessary to bring to light the sparse nature of \eqref{eq_z} recasting the above equation
as a standard sparse model. From an intuitive point of view, note that radar system steers the beam along several directions to cover the surveillance area,
but backscattered echoes and/or interfering signals hit the system from a few directions only.
With this remark in mind, let us sample the angular sector under surveillance and form a discrete and finite set of angles denoted
by $\Theta=\left\{ \theta_1,\ldots, \theta_L \right\}$, $\theta_1\leq\theta_2\leq \ldots \leq \theta_L$.
Moreover, we assume that the target nominal angle $\theta_T$ and the AOA of possible
$N_q\ll L$ coherent interferers belong to $\Theta$. Thus, if we define
$\bV =[\bv(\theta_1),...,\bv(\theta_L)]\in \mathbb{C}^{N\times L}$ as the model matrix whose columns are the steering vectors
associated with the angular positions $\left\{ \theta_1,...,\theta_L \right\}$ and
a vector $\balpha=[\alpha_1,...,\alpha_{L}]^T\in \mathbb{C}^{L\times 1}$ whose
nonzero entries correspond to the AOAs of the target and the coherent interferers in $\bV$, then it is possible to recast $\bz$ as
\be
\ds\bz = \bV \balpha + \bn,
\label{eq_zz}
\ee
where $\balpha$ is assumed to contain the target response $\alpha_T$ as well as the magnitudes of the coherent jammers $\left\{\beta\right\}_{i=1}^{N_q}$.
It is important to observe that since $N_q\ll L$, then $\balpha$ is a sparse vector. In fact, from \eqref{eq_zz}, it turns out that only
$N_q+1$ components of $\balpha$ are possibly different from zero.
In this case, the SLIM
algorithm can be used to produce a very accurate representation
for the scene of interest.
Remarkably, we can exploit the sparse estimate returned by the SLIM to address the following classification problem
\begin{itemize}
\item {\em target plus noise-like interferers hypothesis}:
\be
\cH_1: \ \bz = \alpha_T \bv(\theta_T) + \bn;
\ee
\item {\em noise-like plus coherent interferers hypothesis}:
\be
\cH_2: \ \bz = \sum_{i=1}^{N_{q}}\beta_i \bv(\theta_{q,i}) + \bn;
\ee
\item {\em target plus noise-like and coherent interferers hypothesis}:
\be
\cH_3: \ \bz = \alpha_T \bv(\theta_T) + \sum_{i=1}^{N_{q}}\beta_i \bv(\theta_{q,i}) + \bn.
\ee
\end{itemize}
Thus, as shown in what follows, the newly proposed architecture exhibits, as a byproduct, signal classification capabilities.
Let us start the design by writing the LRT based upon the CUT
\be
\Lambda (\bz;\balpha,\bM_1) = \frac{f_{1}(\bz;\balpha,\bM_1,H_{1,1})}{f_{0}(\bz;\bzero,\bM_1,H_{0,0})} \testp \eta,
\label{eq_glrtM1}
\ee
where $f_{l}(\bz;l \balpha,\bM_1,H_{l,l})$ is the pdf of $\bz$ under $H_{l,l}$, $l=0,1$, whose expression is
\begin{equation}
f_{l}(\bz;l \balpha,\bM_1,H_{l,l}) = \frac{1}{\pi^N \det(\bM_1)}\exp\{ -\tr[\bM_1^{-1}(\bz-l\bV\balpha)(\bz-l\bV\balpha)^\dag]  \}.
\label{eq_pdfz1}
\end{equation}
Now, note that decision rule \eqref{eq_glrtM1} is not of practical interest since both $\balpha$ and $\bM_1$ are not known and, hence, must be estimated
from data. As already stated at the beginning of this subsection, the estimate of $\bM_1$ can be accomplished using the procedures
described in Algorithm \ref{alg:Framwork}. As for $\balpha$, it is estimated resorting to the framework proposed in \cite{slim}. Specifically,
let us assume that $\balpha$ is a random vector independent of the noise component and that obeys a
prior promoting the sparsity, given by
\be
f(\balpha;q) = \frac{1}{C} \prod\limits_{i=1}^{L} \exp{\left \{ -\frac{2}{q} \left( | \alpha_i |^q - 1 \right) \right \} },
\label{eq_prior}
\ee
where $C$ is a normalization constant and  $q\in\Omega_q=(0, 1]$ is a tuning parameter (smaller values of $q$ correspond to sharper peak of the prior distribution and consequently sparser estimate of $\balpha$). Then, $\balpha$ is estimated solving the following maximization problem
\be
\max\limits_{\balpha} f_1(\bz;\widehat{\bM}_1,H_{1,1}|\balpha)f(\balpha;q),
\label{eq_glrt_slim1}
\ee
where $f_1(\bz;\widehat{\bM}_1,H_{1,1}|\balpha)$ is the conditional pdf of $\bz$ given $\balpha$. Taking the negative logarithm, problem
\eqref{eq_glrt_slim1} is equivalent to
\be
\dmin_{\balpha} \underbrace
{\left \{ \| \by - \bA \balpha \|_2^2 + \sum\limits_{i=1}^{L}  \frac{2}{q} \left( | \alpha_i |^q  - 1 \right) \right \}}_{g_q(\balpha)}
\label{eq_glrt_slim2}
\ee
where $\bA=\widehat{\bM}_1^{-1/2}\bV$ and $\by=\widehat{\bM}_1^{-1/2}\bz$.
Notice that the first addendum of $g_q(\balpha)$ corresponds to a fitting term, whereas the second term promotes sparsity.
Setting to zero the first derivative\footnote{We make use of the following definition for the derivative of a
real function $f(\alpha)$ with respect to the complex argument $\alpha=\alpha_r+j \alpha_i$,
$\alpha_r, \alpha_i \in \R$, \cite{VanTrees4}
\be
\frac{\partial f(\alpha)}{\partial \alpha} = \frac{1}{2} \left[ \frac{\partial f(\alpha)}{\partial \alpha_r}
+ j \frac{\partial f(\alpha)}{\partial \alpha_i}\right].
\ee}
of $g_{q}(\balpha)$ with respect to $\balpha$ leads to
\be
\frac{d}{d\balpha}  [g_{q}(\balpha)] = \bA^\dag \bA \balpha - \bA^\dag \by + \bP_q^{-1}\balpha = \bzero,
\label{eq_der0}
\ee
where  $\bP_q = \diag{(\bp_q)}$, with $\bp_q=[|\alpha_1|^{2-q},|\alpha_2|^{2-q},...,|\alpha_L|^{2-q}]^T$.
Supposing that an initial estimate of $\balpha$ is available, it is possible to apply a cyclic optimization procedure as in \cite{slim}, and
the step at the $(m)$th iteration can be expressed as
\be
\label{eq_gamma}
\balpha_q^{(m)} = \bP_q^{(m-1)} \bA^\dag \left( \bA \bP_q^{(m-1)} \bA^\dag +  \bI \right)^{-1} \by,
\ee
given $\bP_q^{(m-1)}=\diag (\bp_q^{(m-1)})$ from the $(m-1)$th iteration.
The optimization procedure can terminate after a fixed number of iterations or
when the following convergence criterion is satisfied
\be
\frac{\| \balpha_q^{(m)} - \balpha_q^{(m-1)} \|_2}{\| \balpha_q^{(m)} \|_2} < \Delta,
\label{eq_stopcriterion}
\ee
with $\Delta$ a suitable small positive number.
As for the initial value of $\balpha$, a possible choice is
\be
\alpha_i^{(0)} = \frac{\bv(\theta_i)^\dag \widehat{\bM}_1^{-1}\bz}{\bv(\theta_i)^\dag \widehat{\bM}_1^{-1}\bv(\theta_i)}, \quad i=1,\ldots,L.
\label{eqn:mle0}
\ee
It still remains to estimate $q\in\Omega_q$. As a preliminary step, we sample $\Omega_q$ to come up with a
finite set of admissible values for $q$ denoted by $\bar{\Omega}_q$.
Now, given $q\in\bar{\Omega}_q$, let $\tilde{\balpha}_{q}$ be the estimate of
$\balpha$ provided by the above iterative procedure, summarized in Algorithm \ref{algSLIM}, and estimate the number of peaks, $h(q)$ say, in $\tilde{\balpha}_{q}$ as follows
\begin{enumerate}
\item sort the entries of $\tilde{\balpha}_{q}$ from the largest to the smallest;
\item select $h(q)$ returning the lowest value of
$\mbox{BIC}_q = 2  \| \by  - {\bA}\widehat{\balpha}_q   \|_2^2  + 3 h(q) \log\left(2N\right)$,
where $3h(q)$ is the number of parameters to be estimated (namely, azimuth and complex amplitude for each active peaks) and $\widehat{\balpha}_{q}$ is the least-squares estimate for the selected peaks setting to zero the other entries of $\balpha$ (denote by $\bar{\balpha}_q$ the final estimate of $\balpha_q$).
\end{enumerate}
As a result, we obtain the set $\{\mbox{BIC}_q: q\in\bar{\Omega}_q\}$ and the estimate of $q$ is obtained as
$\widehat{q} = \arg\min_{q\in\bar{\Omega}_q} \mbox{BIC}_q$.
\begin{algorithm}[tp!]
\caption{Sparse Learning Iterative Minimization (SLIM)}
\label{algSLIM}
\begin{algorithmic}[1]
	\REQUIRE  $\Delta>0$,  $q \in (0,1]$, $\bA$, $\by$
	\ENSURE $\tilde{\balpha}_q$
	\STATE Set $m=0$,  $ \balpha^{(0)} = \left[ \frac{\bv(\theta_1)^\dag \widehat{\bM}_1^{-1}\bz}{\bv(\theta_1)^\dag \widehat{\bM}_1^{-1}\bv(\theta_1)},
	\ldots, \frac{\bv(\theta_L)^\dag \widehat{\bM}_1^{-1}\bz}{\bv(\theta_L)^\dag \widehat{\bM}_1^{-1}\bv(\theta_L)}\right]^T$,
	\STATE Set {$m=m+1$}
	\STATE Compute $\bP_q^{(m-1)}=\diag(\bp_q^{(m-1)})$,
 with $\bp_q^{(m-1)}=\left[ |\alpha_1^{(m-1)}|^{2-q}, |\alpha_2^{(m-1)}|^{2-q}, ..., |\alpha_{L}^{(m-1)}|^{2-q}  \right]^T$
	\STATE Compute $\balpha^{(m)} = \bP_q^{(m-1)} \bA^\dag \left( \bA \bP_q^{(m-1)} \bA^\dag +  \bI \right)^{-1} \by$
	\STATE If ${\| \balpha^{(m)} - \balpha^{(m-1)} \|_2}/{\| \balpha^{(m)} \|_2} < \Delta$ go to step 6 else go to step 2
	\STATE Return $\tilde{\balpha}_q=\balpha^{(m)}$.
\end{algorithmic}
\end{algorithm}
Finally, the adaptive LRT can be written as
\be
\frac{f_{11}(\bz;\bar{\balpha}_{\hat{q}},\widehat{\bM}_1,H_{1,1})}{f_{00}(\bz;\bzero,\widehat{\bM}_1,H_{0,0})} \testp \eta.
\label{eq_glrtbis}
\ee
Before concluding this section, we discuss the classification capabilities raising from $\bar{\balpha}_{\hat{q}}$. 
{ 
Specifically, let us recall that $H_{1,1}$ can be viewed as the union of 
three hypotheses, namely $H_{1,1} = \cup_{i=1}^3 \cH_i$. Now, in order to mitigate
the presence of false objects (ghosts) introduced by $\bar{\balpha}_{\hat{q}}$ and to merge 
contiguous estimates (induced by energy spillover), we partition the set $\Theta$ into $N_s$ subsets, $\Theta_i$ say, 
containing contiguous AOAs. In addition, for simplicity, we assume that $|\Theta|/N_s
=|\Theta_1|=\ldots=|\Theta_{N_s}|=N_{\theta}\in\N$, namely that all the subsets share the same number of contiguous AOAs.
As a consequence, the generic subset can be written 
as 
\be
\Theta_i=\{\theta_{(i-1)N_{\theta}+1},\ldots, \theta_{i N_{\theta}} \}, \ i=1,\ldots,N_s.
\ee
Since the target AOA, $\theta_T$ say, is supposed to belong to $\Theta$, then
there exists $\bar{i}\in\{1,\ldots,N_s\}$ such that $\theta_T\in\Theta_{\bar{i}}$ and let us denote
this subset as $\Theta_T$ ($=\Theta_{\bar{i}}$). 
Thus, for classification purposes, we say 
that a generic subset $\Theta_{i}$, $i\in\{1,\ldots,N_s\}$, contains coherent components if 
there exists at least an index $l\in\{1,\ldots,L\}$ such that 
\be
\bar{\balpha}_{\hat{q}}(l)\neq 0 \quad \mbox{and} \quad \theta_l\in\Theta_{i}.
\label{eqn:conditionPartition}
\ee
The above partitioning procedure allows us to define new vectors, $\bgamma$ and $\bar{\bgamma}_{\hat{q}}$ say, 
of size $N_s$ starting from 
$\balpha$ and $\bar{\balpha}_{\hat{q}}$, respectively, that contain information 
about the angular location in terms of $\Theta_i$ of the
coherent signals received by the system. Such vectors will be used at the analysis stage to quantify the algorithm 
capability in drawing a picture of the entire operating scenario.
Specifically, $\forall i\in\{1,\ldots,N_s\}$, we set 
\begin{itemize}
\item $\bgamma(i)=1$ if condition \eqref{eqn:conditionPartition} applied to $\balpha$ holds, otherwise $\bgamma(i)=0$; 
\item $\bar{\bgamma}_{\hat{q}}(i)=1$ if \eqref{eqn:conditionPartition} is valid, 
otherwise $\bar{\bgamma}_{\hat{q}}(i)=0$.
\end{itemize}
Then, denoting by $\Omega_{\gamma}$, with $|\Omega_{\gamma}|\geq 1$, the set of integers indexing the nonzero entries 
of $\bar{\bgamma}_{\hat{q}}$, we can reason according to the following rationale
\begin{itemize}
\item if data contain only one coherent component ($|\Omega_{\gamma}|=1$),
which can be due to either an interferer or a target, then there exists $\tilde{i}\in\{1,\dots,N_s\}$ 
such that $\Omega_{\gamma}=\{\tilde{i}\}$ (namely, $\bar{\bgamma}_{\hat{q}}(\tilde{i})\neq 0$ 
and $\bar{\bgamma}_{\hat{q}}(i)= 0$, $\forall i\neq\tilde{i}$) and two cases can occur
\begin{itemize}
\item {\em Case 1:} $\Theta_{\tilde{i}}=\Theta_T$, which implies that $\cH_1$ holds true;
\item {\em Case 2:} $\Theta_{\tilde{i}}\neq\Theta_T$, which implies that $\cH_2$ is in force;
\end{itemize}
\item if data contain more than one coherent component ($|\Omega_{\gamma}|>1$), which can be generated
by jammers or the target, then the following cases have to be accounted for 
\begin{itemize}
\item {\em Case 1:} $\exists \tilde{i}\in\Omega_{\gamma}$ such that $\Theta_{\tilde{i}}=\Theta_T$; 
in this case $\cH_3$ is declared;
\item {\em Case 2:} $\forall \tilde{i}\in \Omega_{\gamma}$: $\Theta_{\tilde{i}}\neq\Theta_T$; in this case $\cH_2$ is declared.
\end{itemize}
\end{itemize}
Finally, the SLIM-based detector \eqref{eq_glrtbis} can be incorporated into the architecture depicted in 
Figure \ref{fig_SLIMarchitecture}, where the 
condition on $\tilde{i}$ clearly is: $\exists \ \tilde{i}\in\Omega_{\gamma}: \Theta_{\tilde{i}}=\Theta_T$.
}
It is important to highlight that such architecture can absolve the functions of both SLB and SLC \cite{antennaBased}.
In fact, the use of $\widehat{\bM}_1$ allows to place nulls along the NLJ directions, while
$\bar{\balpha}_{\hat{q}}$ allows to separate the target response from the coherent interferers.

Summarizing, the proposed approach allows to suitably handle situations where NLJs as well as CJs attack the victim radar providing
a tool for the discrimination between useful structured returns and unwanted signals. In fact, focusing on the CUT only, the actual classification problem
herein addressed is the following multiple-hypothesis test
\begin{equation}
\label{eq_finalprob}
\left\{
\begin{array}{l}
\cH_1: \bz = \alpha_T \bv(\theta_T) + \bn,
\\
\cH_2: \bz = \sum\limits_{i=1}^{N_{q}}\beta_i \bv(\theta_{q,i}) + \bn,
\\
\cH_3: \bz = \alpha_T \bv(\theta_T) + \sum\limits_{i=1}^{N_{q}}\beta_i \bv(\theta_{q,i}) + \bn,
\\
H_{00}: \bz = \bn.
\end{array}
\right.
\end{equation}

\section{Illustrative Examples}
\label{Sec:Examples}
In this section, we analyze the performance of the new ECCM strategies against the disturbance injected by NLJs and/or CJs through the antenna sidelobes.
Specifically, in Subsection \ref{NLJ-only}, we analyze the performance in the first scenario (NLJ-only attack) while
in Subsection \ref{SLIM-based}, we investigate the behavior of the SLIM-based detector
when both NLJs and CJs are present.

\subsection{NLJ-only Case}
\label{NLJ-only}
In this section, we present illustrative examples assessing the performance of the multi-stage architectures
devised in Section \ref{Sec:NLJ-only} in terms of $P_d$ against the SINR.
For comparison purpose, we also report the $P_d$ curves of the so-called Double Trained-AMF (DT-AMF) introduced in \cite{doubleTraining},
the IDT-AMF with known $r$, and the clairvoyant (non-adaptive) detector, i.e., the Matched Filter (MF) with known $\bM_1$, which represents
an upper bound to the detection performance. The numerical examples are obtained resorting to standard Monte Carlo counting techniques. More precisely,
the $P_d$ is estimated over $10^3$ independent trials, whereas the detection thresholds are computed exploiting $100/P_{fa}$ independent trials.
In all the illustrative examples, we set $N=16$, $\sigma^2=1$, $d=\lambda/2$, and $P_{fa} = 10^{-4}$. The SINR is defined as
$\textrm{SINR}=|\alpha|^2\bv(0)^{\mathrm{\dag}}\bM_1^{-1}\bv(0)$,
while the considered scenario comprises three NLJs with the same power from
the following AOAs: $\theta_{nj,1}=15^{\circ}$, $\theta_{nj,2}=25^{\circ}$, and $\theta_{nj,3}=-10^{\circ}$.
Then, the resulting ICMs are given by
$\bM_2=\bI+\sum_{i=1}^{3}\textrm{JNR}\ \bv(\theta_{nj,i})\bv(\theta_{nj,i})^{\dag}$ and $\bM_1=\bM_2+\textrm{CNR}\ \bM_c$.
The $(i,j)$th entry of $\bM_c$ is given by $\bM_c(i,j)=\rho^{|i-j|}$, where $\rho=0.9$ is the one-lag correlation coefficient. Finally, the maximum number of NLJs is set to $N/2$ and the GIC parameter, $\rho$ say, is equal
to 2 (this choice represents a reasonable compromise to limit the model overestimation).

\begin{figure}[htp!]
    \centering
    \includegraphics[width=.55\textwidth]{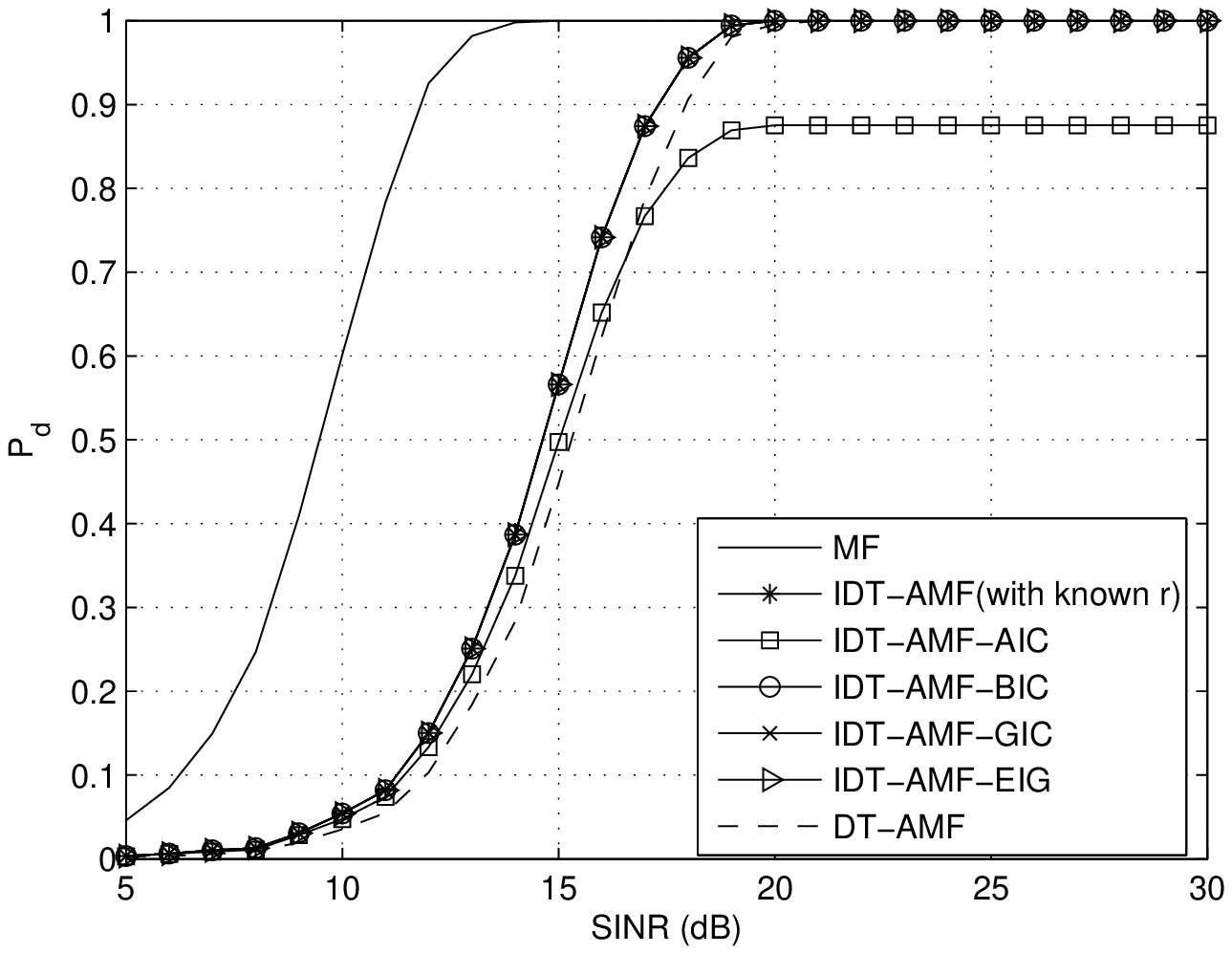}
    \caption{\textit{$P_d$ versus SINR for the MF, IDT-AMF, IDT-AMF-AIC, IDT-AMF-BIC, IDT-AMF-GIC, IDT-AMF-EIG and DT-AMF assuming $N=16$, $K=M=20$, CNR=20 dB and JNR=30 dB. }}
    \label{p1}
\end{figure}

In Figure \ref{p1}, the $P_d$ versus SINR for all the considered detectors is plotted assuming $K = M = 20$, JNR $= 30$ dB and CNR $= 20$ dB.
As it can be seen, the IDT-AMF-BIC, IDT-AMF-GIC and IDT-AMF-EIG have nearly the same performance as the IDT-AMF with known $r$ and
they exhibit higher $P_d$ values than the DT-AMF with a gain of 0.6 dB at $P_d=0.9$.
The IDT-AMF-BIC, IDT-AMF-GIC, IDT-AMF-EIG and IDT-AMF exhibit similar performances due
to the fact that the stages responsible for the estimate of $r$ share the same estimation accuracy.
As to the IDT-AMF-AIC, it experiences a loss about 1.0 dB at $P_d=0.8$ with respect to other proposed detectors but still has
slightly higher $P_d$ than the DT-AMF in the low/medium SINR region. However, the IDT-AMF-AIC is not capable of
achieving $P_d=1.0$ for higher SINR values at least for the considered parameter values.

\begin{figure}[htp!]
    \centering
    \includegraphics[width=.55\textwidth]{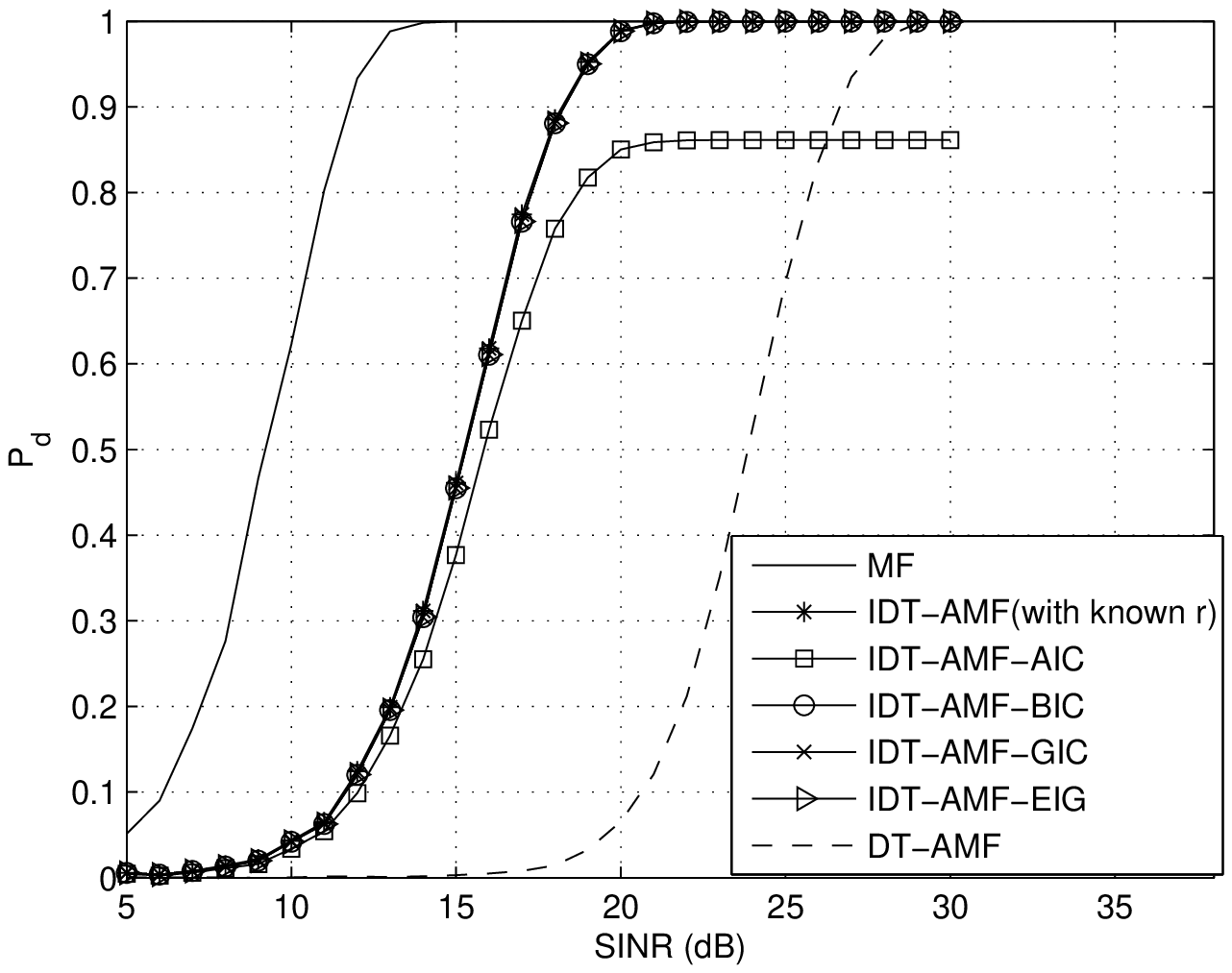}
    \caption{\textit{$P_d$ versus SINR for the MF, IDT-AMF, IDT-AMF-AIC, IDT-AMF-BIC, IDT-AMF-GIC, IDT-AMF-EIG and DT-AMF assuming $N=16$, $K=14$, $M=20$, CNR=20 dB and JNR=30 dB. }}
    \label{p2}
\end{figure}

\begin{figure}[htp!]
    \centering
    \includegraphics[width=.55\textwidth]{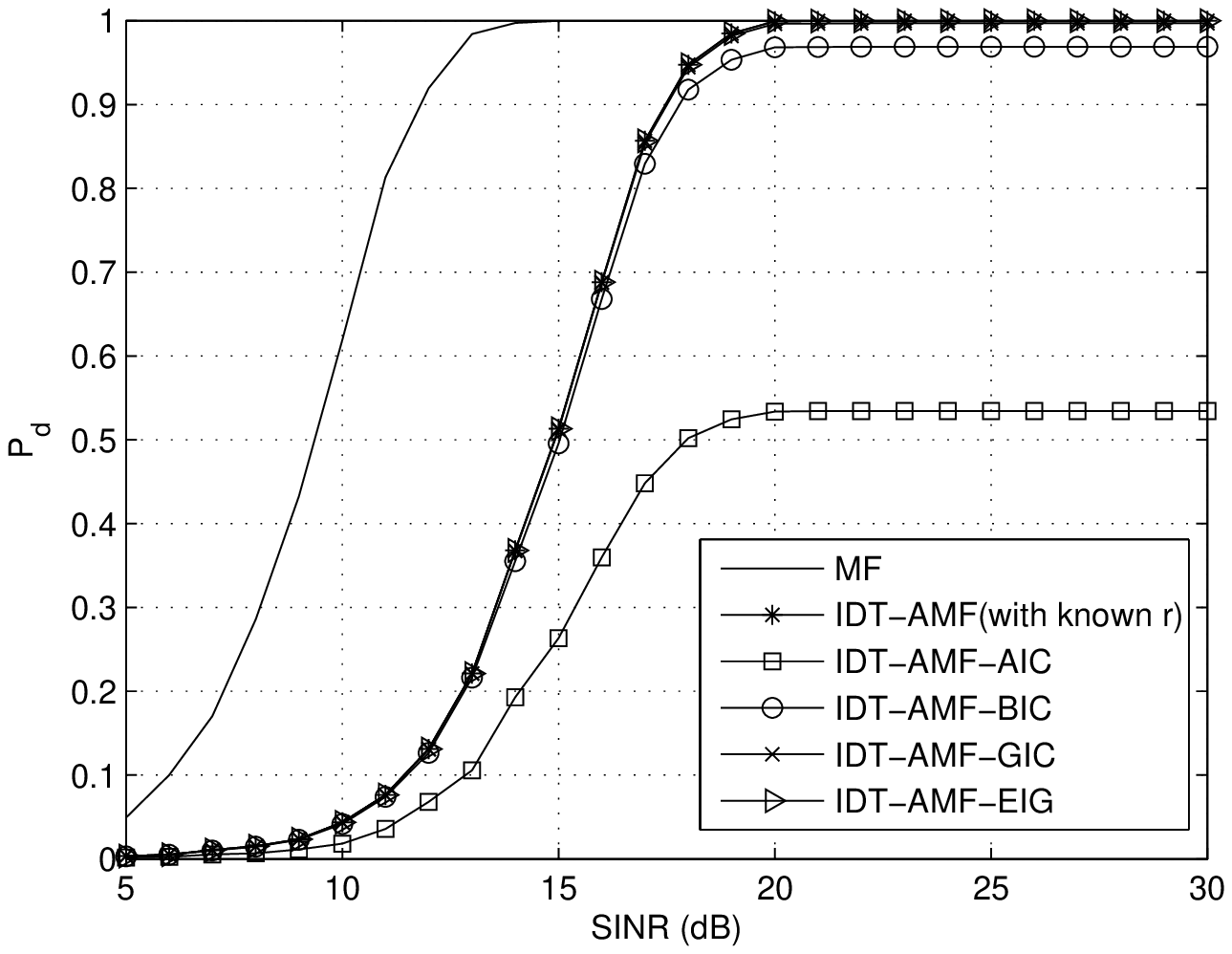}
    \caption{\textit{$P_d$ versus SINR for the MF, IDT-AMF, IDT-AMF-AIC, IDT-AMF-BIC, IDT-AMF-GIC, and IDT-AMF-EIG  assuming $N=16$, $K=20$, $M=13$, CNR=20 dB and JNR=30 dB. }}
    \label{p3}
\end{figure}

To show the influence of $K$ and $M$ on the detection performance of the proposed detectors, in Figure \ref{p2}
we set $K=14$ leaving the other parameters as in Figure \ref{p1}, whereas the parameter values in Figure \ref{p3} are the same as in Figure \ref{p1}
but for $M = 13$. Inspection of Figure \ref{p2} confirms the trend observed in Figure \ref{p1}. Moreover, the performance gain of the proposed detectors
with respect to the DT-AMF increases as $K$ decreases. Precisely, the DT-AMF experiences a loss of about 8.5 dB at $P_d=0.9$ with respect to
the architectures based upon BIC, GIC, and the modified ML estimation. Even though the IDT-AMF-AIC performs better than the DT-AMF for SINR$<26$ dB,
it is still not capable of ensuring $P_d=1$. Comparing Figure \ref{p2} with Figure \ref{p1}, we can note that each proposed detector experiences a
loss of about 1 dB when $K$ decreases from $20$ to $14$. This is due to the fact that the estimation quality of $r$ and $\bM_1$ reduces.
On the other hand, Figure \ref{p3} highlights that, when $M=13$ and $K=20$, the IDT-AMF-GIC and IDT-AMF-EIG overcome
the IDT-AMF-BIC with a gain of 0.4 dB
at $P_d=0.9$, whereas the IDT-AMF-AIC has a severe performance degradation. It is important to stress that the $P_d$ curve of DT-AMF is
not reported in Figure \ref{p3} since it is not defined when $M<N$.

\begin{figure}[htp!]
    \centering
    \includegraphics[width=.55\textwidth]{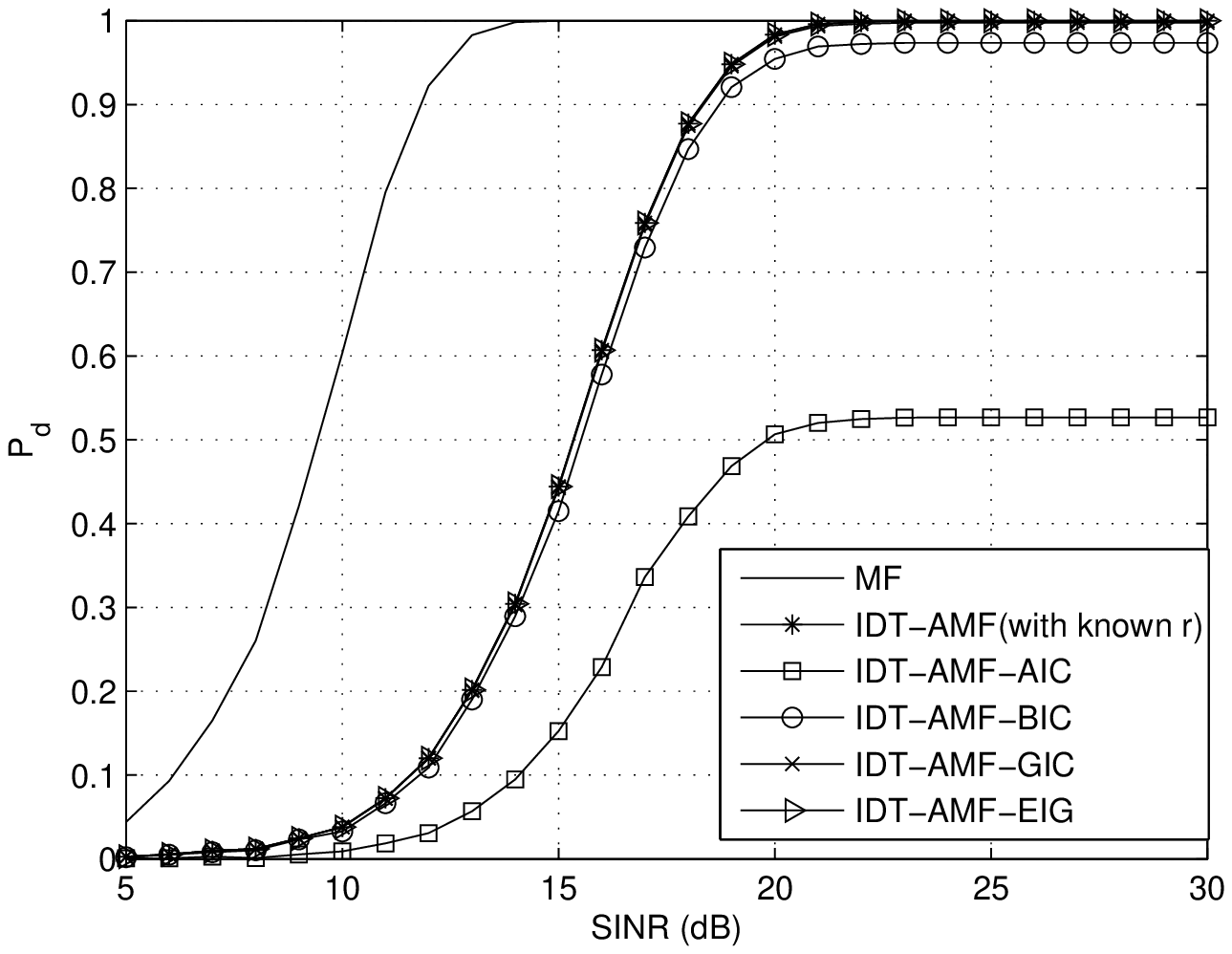}
    \caption{\textit{$P_d$ versus SINR for the MF, IDT-AMF, IDT-AMF-AIC, IDT-AMF-BIC, IDT-AMF-GIC and IDT-AMF-EIG assuming $N=16$, $K=14$, $M=13$, CNR=20 dB and JNR=30 dB. }}
    \label{p4}
\end{figure}

Finally, in Figure \ref{p4}, the $P_d$ performances are investigated assuming $K < N$ and $M < N$. In particular, we set $K = 14$, $M = 13 $
and leave unaltered the other parameters. The curves in Figure \ref{p4} indicate that the detection performances of the
IDT-AMF-GIC, IDT-AMF-EIG, and IDT-AMF are still similar while the IDT-AMF-BIC experience a performance degradation of about 0.5 dB at $P_d=0.9$.

Summarizing, architectures IDT-AMF-GIC, IDT-AMF-EIG, and IDT-AMF-BIC are effective solutions to detect point-like targets in
the presence of an unknown number of NLJs, with IDT-AMF-GIC and IDT-AMF-EIG slightly superior to IDT-AMF-BIC. However,
IDT-AMF-GIC requires to set a parameter while IDT-AMF-EIG exploits two thresholding stages. For this reason, the IDT-AMF-BIC emerges
as a viable means for practical implementation.

\subsection{SLIM-based detector performance analysis (NLJs and CJs joint attack)}
\label{SLIM-based}
In this subsection, we investigate the behavior of the SLIM-based detector in a scenario which assumes the joint presence
of one NLJ and two CJs ($N_q=2$). It is important to note that CJs can be also categorized as targets, since they emulate echoes from an object
of interest. For this reason, the considered performance metrics concern the capability of the system to detect both target and CJs and to discriminate
between the echoes backscattered from the target and the echo-like signal transmitted by the CJs.
The NLJ illuminates the radar with a JNR of $30$ dB and AOA $\theta_{nj}=10^\circ$, whereas
the CJs are located at $\theta_{q,1}=-14^\circ$ and $\theta_{q,2}=16^\circ$ and radiate power at the same JNR of 45 dB.
Target signature is given by $\bv(0)$ and the SINR is defined as in the previous subsection.
In other words, the operating scenario corresponds to $\cH_3$.
As for the ICMs, $\bM_2=\bI+\mbox{JNR} \bv(\theta_{nj})\bv(\theta_{nj})^\dag$, whereas $\bM_1$ is defined using the
same parameters as in the previous subsection.

The analysis, conducted by means of Monte Carlo simulation, is aimed at estimating the following main performance metrics:
\begin{itemize}
\item the probability of detection ($P_d$) defined as the probability to declare $H_{1,1}$ when the latter holds true for a preassigned
value of the $P_{fa}$, defined as the probability to declare $H_{1,1}$ when $H_{0,0}$ is in force;
\item the probability of declaring the presence of a target under $\cH_3$, which is denoted by $P_{t|H_3}$;
\item the probabilities of correct classification, namely the probability of declaring $\cH_i$, $i=1,2,3$, when it is on force.
\end{itemize}
Finally, to assess the estimation capabilities of the SLIM-based detector, additional figure of merits
will be suitably introduced in the second part of this section.
All the mentioned metrics are estimated resorting to $10^3$ independent trials, while the detection threshold is computed over $100/P_{fa}$ independent trials.
An additional thresholding of the entries of $\bar{\balpha}_{\hat{q}}$ is applied to mitigate as much as possible the number of
false targets generated by the SLIM estimate especially at low SINR values. To this end, the threshold is set to
ensure a probability of declaring the presence of a false target equal to 10$^{-2}$.
Finally, all the numerical examples assume $N=16$, $P_{fa}=10^{-4}$, an angular sector under surveillance ranging from $-22^\circ$ to $22^\circ$
and uniformly sampled at $1$ degree (namely, $L=45$), and $|\Theta_i|=5$ with $\Theta_T=\{-2,\ldots,2\}$.

\begin{figure}
    \centering
    \includegraphics[width=0.55\textwidth]{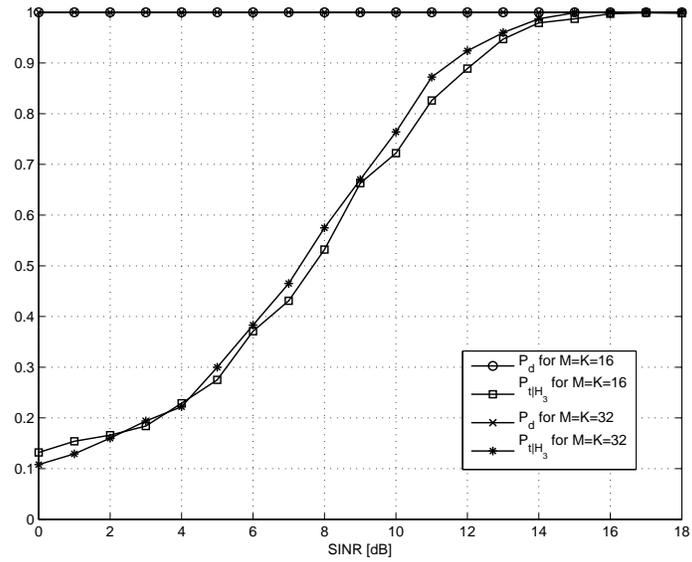}
    \caption{Performance of the SLIM-based detector in terms of $P_{d}$ and $P_{t|H_3}$ versus SINR for $M=K=16$ and $M=K=32$.}
    \label{fig:my_label}\label{fig_perf_slim}
\end{figure}

In Figure \ref{fig_perf_slim}, we show the $P_d$ and the $P_{t|H_3}$ both as functions of the SINR and for $(K,M)\in\{ (16,16), (32,32) \}$.
As expected, the $P_d$ is equal to $1$ regardless the values of $K$, $M$, and SINR. This is due to the presence of the CJs whose JNR is constant and equal
to $45$ dB. On the other hand, the $P_{t|H_3}$ achieves $1$ at SINR$=15$ dB when $(K,M)=(32,32)$ and at SINR$=17$ dB when $(K,M)=(16,16)$.
Generally speaking, inspection of the figure highlights that increasing the volume of training samples leads to a moderate improvement of the $P_{t|H_3}$.

It is important to highlight that the SLIM-based detector draws, as a byproduct, a picture of the electromagnetic scenario under surveillance in terms of
AOAs of possible passive or active objects. However, this picture might contain false objects (ghosts) or
ignore existing sources. Thus, it is worth to evaluate to what extent the above phenomena take place. To this end,
in Figure \ref{fig:metrics}, we plot the following figures of merit as functions of the SINR
\begin{itemize}
\item Root Mean Square (RMS) number of missed interferers, $n_{mj}$ say, evaluated by verifying that the $\Theta_i$s
corresponding to the two jammers refer to null entries of $\bar{\balpha}_{\hat{q}}$;
\item RMS number of ghosts, $n_{g}$ say, defined as the nonzero components of $\bar{\bgamma}_{\hat{q}}$ in positions different from that of the target and CJs;
\item the Hausdorff metric \cite{4567674} between $\bgamma$ and $\bar{\bgamma}_{\hat{q}}$. This metric belongs to the family of the multi-object distances
which are able to capture the error between two sets of vectors and is defined as
$h_d(\cX,\cY) = \max \{ \max_{x \in \cX} \min_{y\in \cY} d(x,y) , \max_{y\in \cY} \min_{x\in \cX} d(x,y)  \}$
with $\cX$ and $\cY$ are the sets of the coordinates of the nonzero entries of $\bgamma$ and $\bar{\bgamma}_{\hat{q}}$, respectively.
\end{itemize}
Note that the Hausdorff metric decreases as the SINR increases up to $15$ dB and then it takes on a constant value equal to $n_{g}=0.4$.
Remarkably, the RMS number of missed jammers is close to zero regardless of the SINR, since it depends on the JNR.
Finally, Figure \ref{fig:hist} contains the classification histograms assuming SINR$=20$ dB.
More precisely, each subplot presents the probabilities $P(\cH_i|\cH_k)$ as the percentages of declaring $\cH_i$, $i=1,2,3$, when $\cH_k$, $k=1,2,3$, is in force.
The histograms highlight that the probability of correct classification, namely of deciding for $\cH_i$ when the latter holds, is
close to $1$ at least for the considered parameter setting.

Summarizing, the analysis shows that the SLIM-based detector is very versatile, since it can operates in the presence of NLJ and/or CJs. More importantly, it
can ensure excellent signal classification performances allowing for the discrimination between the echoes backscattered
from a target and coherent signals emitted by hostile platforms.

\section{Conclusions}
\label{Sec:Conclusions}
In this paper, we have devised adaptive detection architectures with signal-processing-related ECCM capabilities against
the attack of NLJs and/or CJs from the antenna sidelobes.
We have analyzed two operating scenarios which differ for the presence of an unknown number of CJs assuming that
two independent sets of training samples are available for estimation purposes. Next, we have devised novel signal processing procedures to estimate
the ICM capable of providing reliable estimates even in the presence of a low volume of secondary data. Moreover, such estimation procedures
work without knowing the actual number of NLJs. In the case where CJs are present, we have conceived a multistage architecture which
leverages the hidden sparse nature of the data model to detect structured signals backscattered from a target or generated by CJs.
To this end, we have borrowed the SLIM paradigm proposed in \cite{slim}. The performance analyses has highlighted that the newly proposed
detection architectures exhibit satisfactory performances and, more important, the SLIM-based detector with its classification capabilities
can act as an improved SLB, since, in the case where a target and CJs are simultaneously present, it does not blank the possible detection.

Future research tracks might encompass the design of detection architectures for range-spread targets based upon compressive sensing algorithms.

\section{ACKNOWLEDGMENTS}
This work was partially supported by the National Natural Science Foundation of China under Grant No. 61571434, and Chinese academy of sciences president's international fellowship initiative under Grant No. 2018VTB0006.

\begin{appendices}

\section{MLE of $\bM_2$ for known $r$}
\label{app:M2_estimate}

In this Appendix, we provide the derivation of \eqref{eqn:M2_estimate}.
To this end, compute the logarithm of $\cL_R(\sigma^2,\bM_{nj})$ and recast it
by means of the eigendecompositions of $\bM_2$ and $\bR\bR^\dag$ as
\begin{align}
\ln \cL(\sigma^2,\bM_{nj}) &=-MN\ln\pi-M\left\{\sum_{i=1}^{r}\ln(\sigma^2+\lambda_{nj,i})+(N-r)\ln\sigma^2\right\} \nonumber
\\
&  -\mathrm{tr}\left[(\sigma^2\textbf{I}+\bLambda_{nj})^{-1}\bU^{\dag}\bU_{S1}\bLambda_{S1}\bU_{S1}^{\dag}\bU\right] \nonumber
\\
& = h(\sigma^2,\bLambda_j,\bU),
\end{align}
where
\begin{itemize}
\item $\bLambda_{nj}\in \R^{N\times N}$ is a diagonal matrix whose nonzero entries are the eigenvalues
of $\bM_{nj}$ with $\lambda_{nj,1}\geq\lambda_{nj,2}\geq...\geq\lambda_{nj,r}>0$ and $\bU \in \C^{N\times N}$ is the unitary matrix of
the corresponding eigenvectors;
\item $\bLambda_{S1} \in \R^{N\times N}$ is a diagonal matrix whose nonzero entries are the eigenvalues of $\bR\bR^\dag$, denoted by $\gamma_{1,1}\geq \gamma_{1,2}\geq...\geq\gamma_{1,N}\geq 0$, and $\bU_{S1} \in \C^{N\times N}$ contains the corresponding eigenvectors.
\end{itemize}
Thus, the maximization of $\ln \cL(\sigma^2,\bM_{nj})$ with respect to $\bM_2$ is equivalent to
\be\label{eqn:16}
\underset{\sigma^2,\bLambda_j,\bU}{\max}h(\sigma^2,\bLambda_j,\bU).
\ee
Now, the optimization with respect to $\bU$ can be accomplished exploiting \emph{Theorem 1} \cite{mirsky1959trace}, we obtain that
\begin{align}
\max\limits_{\bU} \mathrm{tr}\left[(\sigma^2\textbf{I}+\bLambda_j)^{-1}\bU^{\dag}\bU_{S1}\bLambda_{S1}\bU_{S1}^{\dag}\bU\right]
&=\max\limits_{\bW_1}\mathrm{tr}[(\sigma^2\textbf{I}+\bLambda_j)^{-1}\bW_1\bLambda_{S1}\bW_1^{\dag}]\nonumber\\
&=\mathrm{tr}[(\sigma^2\textbf{I}+\bLambda_j)^{-1}\bLambda_{S1}],
\end{align}
where $\bW_1=\bU^{\dag}\bU_{S1}$. It is possible to show that optimization with respect to $\bW_1$ leads to $\bW_1=\bI e^{j\theta_1}$ for arbitrary $\theta_1\in [0,2\pi]$. Thus, choosing for simplicity $\theta_1=0$, an MLE of $\bU$ can be recast as $\widehat{\bU}=\bU_{S1}$.
As a consequence, problem \eqref{eqn:16} becomes
\be
\underset{\sigma^2,\bLambda_j}{\max} g(\sigma^2,\bLambda_j),
\ee
where
\be
g(\sigma^2,\bLambda_j)=-MN\ln \pi-M\left\{\sum_{i=1}^{r}\ln(\sigma^2+\lambda_{j,i})+(N-r)\ln\sigma^2\right\}-\sum_{i=1}^{r}\frac{\gamma_{1,i}}{\sigma^2+\lambda_{j,i}}-\frac{1}{\sigma^2}\sum_{i=r+1}^{N}\gamma_{1,i}.
\ee
To estimate the remaining parameters, let us set to zero the gradient of $g(\sigma^2,\bLambda_j)$. Then, the resulting estimates are given by
\be
\widehat{\sigma}^2 =\frac{1}{M(N-r)}\sum\limits_{i=r+1}^{N}\gamma_{1,i} \quad \mbox{and} \quad
\widehat{\lambda}_{j,i}+\widehat{\sigma}^2 = \frac{\gamma_{1,i}}{M}, \quad i=1,...,r.
\ee
Finally, the MLE of $\widehat{\bM}_2$ is
\be\label{eq15}
\widehat{\bM}_2=\bU_{S1}(\widehat{\sigma}^2\bI+\widehat{\bLambda}_j)\bU_{S1}^{\dag},
\ee
where $\widehat{\sigma}^2\bI+\widehat{\bLambda}_j=\textbf{diag}\left\{\frac{\gamma_{1,1}}{M},\ldots,\frac{\gamma_{1,r}}{M},
\frac{1}{M(N-r)}\sum_{i=r+1}^{N}\gamma_{1,i},\ldots,\frac{1}{M(N-r)}\sum_{i=r+1}^{N}\gamma_{1,i}\right\}$.

\section{MLE of $\bM_c$ for known $\bM_2$}
\label{app:Mc_estimate}
Assume that $\bM_2$ is known and consider the following maximization problem
\be
\dmax_{\bM_c} \cL_{Z}(\bM_c),
\label{eqn:estimationMc01}
\ee
where $\cL_{Z}(\bM_c)$ is defined by \eqref{eqn:L_Zdef}. To solve \eqref{eqn:estimationMc01}, let us recast the logarithm of $\cL_{Z}(\bM_c)$ as
\begin{align}\label{eqn:17}
&-KN\ln\pi-K\ln\det(\bM_2)-K\ln\det(\bI+\bM_2^{-\frac{1}{2}}\bM_c\bM_2^{-\frac{1}{2}}) \nonumber
\\
&-\mathrm{tr}\left[(\bI+\bM_2^{-\frac{1}{2}}\bM_c\bM_2^{-\frac{1}{2}})^{-1}\bM_2^{-\frac{1}{2}}\bZ\bZ^\dag\bM_2^{-\frac{1}{2}}\right] \nonumber
\\
&=-KN\ln\pi-K\ln\det(\bM_2)-K\ln\det(\bI+\bOmega_c)-\mathrm{tr}\left[(\bI+\bOmega_c)^{-1}\bV^{\dag}\bU_{S2}\bLambda_{S2}\bU_{S2}^{\dag}\bV\right] \nonumber
\\
&=\cL_{Z}(\bV,\bOmega_c),
\end{align}
where the last equality is due to the eigendecomposition of $\bM_2^{-\frac{1}{2}}\bM_c\bM_2^{-\frac{1}{2}}$ and $\bZ\bZ^\dag$. In fact,
in \eqref{eqn:17}, $\bLambda_{S2} \in \R^{N\times N}$ is a diagonal matrix whose nonzero entries are the eigenvalues of $\bM_2^{-\frac{1}{2}}\bZ\bZ^\dag\bM_2^{-\frac{1}{2}}$ denoted by $\gamma_{2,1}\geq...\geq\gamma_{2,N}\geq0$ with $\bU_{S2} \in \C^{N\times N}$
the unitary matrix of the corresponding eigenvectors and $\bOmega_c \in \R^{N\times N}$ is the diagonal matrix containing the eigenvalues
of $\bM_2^{-\frac{1}{2}}\bM_c\bM_2^{-\frac{1}{2}}$ denoted by $\lambda_{c,1}\geq...\geq\lambda_{c,N}>0$ with $\bV \in \C^{N\times N}$ the unitary matrix of the corresponding eigenvectors. It follows that problem \eqref{eqn:estimationMc01} becomes
\be
\max_{\bV,\bOmega_c} \cL_{Z}(\bV,\bOmega_c).
\ee
The optimization with respect to $\bV$ can be accomplished adopting the same line of reasoning as for $\bW_1$ in Appendix \ref{app:M2_estimate}, namely
\begin{align}
\max\limits_{\bV}\mathrm{tr}\left[(\bI+\bOmega_c)^{-1}\bV^{\dag}\bU_{S2}\bLambda_{S2}\bU_{S2}^{\dag}\bV\right]&=\max\limits_{\bW_2}\mathrm{tr}[(\bI+\bOmega_c)^{-1}\bW_2\bLambda_{S2}\bW_2^{\dag}]\nonumber
\\
&=\mathrm{tr}[(\bI+\bOmega_c)^{-1}\bLambda_{S2}],
\end{align}
where $\bW_2=\bV^{\dag}\bU_{S2}$ and the last equality comes from the fact that $\widehat{\bW}_2=\bI e^{j\theta_2}$
with $\theta_2 \in [0,2\pi]$ arbitrary. As a result, an estimate of $\bV$ is $\widehat{\bV}=\bU_{S2}$.

The final step consists in solving
\be
\dmax_{\bOmega_c} \bar{g}(\bOmega_c),
\ee
where
\be
\bar{g}(\bOmega_c)=-K\sum_{i=1}^{N}\ln(1+\lambda_{c,i})-\sum_{i=1}^{N}\frac{\gamma_{2,i}}{1+\lambda_{c,i}}.
\ee
Thus, setting to zero the gradient of $\bar{g}(\bOmega_c)$, we obtain
\be
\widehat{\lambda}_{c,i}=
\max\left\{ \frac{\gamma_{2,i}}{K}-1, 0 \right\}, \quad i=1,\ldots,N.
\ee
Gathering the above results, the MLE of $\bM_c$ for known $\bM_2$ is
\be
\widehat{\bM}_c[\bM_2]=\bM_2^{\frac{1}{2}}\bU_{S2}\widehat{\bOmega}_c\bU_{S2}^{\dag}\bM_2^{\frac{1}{2}},
\ee
where $\widehat{\bOmega}_{c}=\textbf{diag}\left\{\widehat{\lambda}_{c,1},\ldots,\widehat{\lambda}_{c,N}\right\}$.
\end{appendices}
\bibliographystyle{IEEEtran}
\bibliography{group_bib}
\begin{table}
\caption{List of Acronyms}
\begin{center}
\begin{tabular}{c|l}
\hline 
\hline
\multicolumn{2}{c}{ Acronyms}\\
\hline 
\hline
AIC & Akaike Information Criterion\\
\hline
AOA & Angle of Arrival\\
\hline
AMF & Adaptive Matched Filter\\
\hline
BIC & Bayesian Information Criterion\\
\hline
CJ & Coherent Jammer\\
\hline
CNR & Clutter-to-Noise Ratio\\
\hline
CUT & Cell Under Test\\
\hline
DT & Double Trained\\
\hline
ECCM & Electronic Counter-countermeasure\\
\hline
ECM & Electronic Countermeasure\\
\hline
GIC & Generalized Information Criterion\\
\hline
GLRT & Generalized Likelihood Ratio Test\\
\hline
ICM & Interference Covariance Matrix\\
\hline
IDT & Improved Double Trained\\
\hline
JNR & Jammer-to-Noise Ratio\\
\hline
LRT & Likelihood Ratio Test\\
\hline
MLA & Maximum Likelihood Approach\\
\hline
MLE & Maximum Likelihood Estimate\\
\hline
NLJ & Noise Like Jammer\\
\hline
MOS & Model Order Selection\\
\hline
RMS & Root Mean Square\\
\hline
SINR & Signal-to-Interference plus Noise Ratio\\
\hline
SLB & Sidelobe Blanker\\
\hline
SLC & Sidelobe Canceler\\
\hline
SLIM & Sparse Learning via Iterative Minimization\\
\hline 
\hline 
\end{tabular}
\end{center}
\label{tab:Acronyms}
\end{table}
\begin{figure}
    \centering
    \includegraphics[width=.6\textwidth]{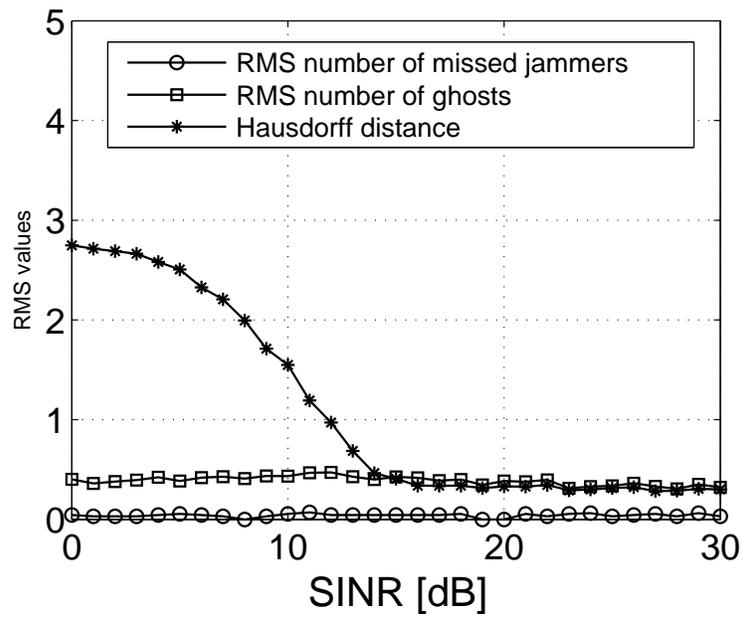}
    \caption{RMS values of $n_{mj}$, $n_{g}$, and $h_d(\cdot,\cdot)$ assuming $M=K=16$.}
    \label{fig:metrics}
\end{figure}
\begin{figure}
    \centering
    \includegraphics[width=.55\textwidth]{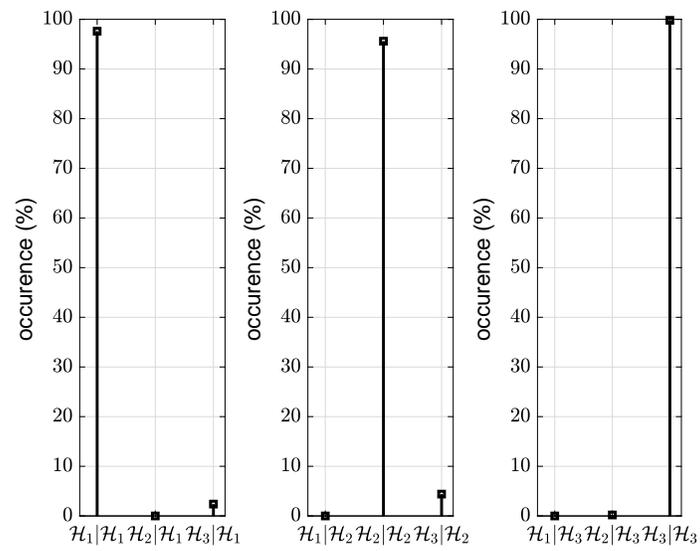}
    \caption{Classification histograms for all the three hypotheses $\mathcal H_{1}$, $\mathcal H_{2}$, and $\mathcal H_{3}$ assuming $M=K=16$ and SINR$=20$ dB.}
    \label{fig:hist}
\end{figure}
\end{document}